\documentclass[twocolumn,showpacs,nofootinbib,preprintnumbers,amsmath,amssymb]{revtex4}
\usepackage{graphicx}
\usepackage{epsfig}
\usepackage{amssymb}
\usepackage{dcolumn}
\usepackage{bm}
\usepackage[hypertex]{hyperref}
\usepackage{psfrag}

\begin{document}

\title{Quasi-stationary states in the self-gravitating sheet model}

\author{Michael Joyce$^{1,2}$ and  Tirawut Worrakitpoonpon$^{1}$}

\affiliation{$^1$Laboratoire de Physique Nucl\'eaire et Hautes \'Energies,\\ 
Universit\'e Pierre et Marie Curie - Paris 6,
CNRS IN2P3 UMR 7585,  4 Place Jussieu, 75752 Paris Cedex 05, France}
\affiliation{$^2$Laboratoire de Physique Th\'eorique de la Mati\`ere Condens\'ee,\\
Universit\'e Pierre et Marie Curie - Paris 6,
CNRS UMR 7600, 4 Place Jussieu, 75752 Paris Cedex 05, France}

\begin{abstract}   
We study quasi-stationary states (QSS) resulting from violent relaxation in 
the one-dimensional self-gravitating ``sheet model'',  revisiting in particular
the question of
the adequacy of the theory of Lynden-Bell (LB)  to describe them.
For ``waterbag'' initial conditions characterized by a single phase space density, the
prediction of this theory is, in this model, a function of only one parameter, which 
can  conveniently be chosen to be the ratio of the energy to that in the 
degenerate limit.  Studying a class of such initial conditions in which
the shape of the initial waterbag is varied, we find that the LB predictions
are reasonably good always in the low energy region, while at higher energies (i.e. in the 
non-degenerate limit) they are  generally not even qualitatively correct, although 
certain initial conditions can still be found where they are as good as at low
energy. We find notably that, in line with what has been observed by Levin et al. in some other 
models, when LB theory does not work the QSS are always characterized by the 
presence of a {\it degenerate} core, which these authors explain as the result of 
dynamical resonances. In short LB theory appears to be a good approximation
only when violent relaxation is sufficiently ``gentle", and otherwise a degenerate
core-halo structure results.  

\end{abstract}


\maketitle

\section{Introduction} \label{intro}
The rich statistical mechanics of long-range interacting systems has been a subject of active 
study in recent years (for a recent review see e.g. \cite{campa_etal_LRreview_2009}). As for self-gravitating systems, such 
systems have been
understood to give rise generically to non-equilibrium states which evolve only on 
time-scales which diverge with the number of particles. The degree to which such 
``quasi-stationary'' states (QSS) can be understood, and their properties predicted, by a statistical 
approach is a question which is inevitably posed.  In this context a theory originally
formulated by Lynden-Bell in the astrophysical context in the sixties \cite{lynden_bell_1967},
and which has been applied also in the study of two dimensional vortices \cite{chavanis_etal_1996}, 
has seen revived interest in recent years.  Study notably of a one-dimensional (1D) toy 
model, the  Hamiltonian Mean Field (HMF) Model which describes particles on a ring interacting
by a cosine potential, showed that this theory can predict sometimes very accurately
the properties of these states (see, e.g., \cite{antoniazzi+fanelli+barre+chavanis+dauxois+ruffo_2007, stanascia_etal_2009}
and references therein),  and more generally manages to capture the qualitative dependence of the QSS on initial conditions. 
While it is clear that  the ``LB theory'' is not entirely adequate in general, these studies suggest that the basic physical principle behind 
it --- maximization of an entropy subjected to the  constraints appropriate to Vlasov dynamics --- 
is, at the very least, a reference point for understanding the out of equilibrium dynamics of these systems.
This contrasts strongly with the view of this theory in the (original) context of the astrophysical 
literature, where it has simply been discarded as a completely inadequate, and basically
irrelevant, theory \cite{tremaine+henon+lynden_bell_1986, arad+lynden_bell_2005}.
One  recent study \cite{levin+pakter+rizzato_2008} of  three 
dimensional (3D) self-gravitating systems concludes, however, that LB theory may indeed be 
relevant also to this case. This study shows that in a certain limited range of initial conditions 
the LB theory predicts well the density profiles of QSS, and proposes an alternative theory
to explain their properties in the regime where LB no longer works well. The same
authors have shown that the same statements apply both to plasma systems \cite{levin+pakter+teles_2008} and
two-dimensional self-gravitating system \cite{teles_etal_2010}, and, in a very recent 
article \cite{pakter+levin_2010}, have used the alternative theory to account for 
QSS in the HMF model. 

In this paper we study these issues in 
the so-called self gravitating ``sheet model''  (SGS) of particles in one dimension attracted by
forces independent of separation. Our main goal is to characterize more precisely than
has been done previously the degree of validity of the LB theory in this model, which
is one of the canonical toy models for the study of such systems, and to determine whether
the properties of the QSS can be characterized in a simple manner and perhaps understood 
when the LB theory does not apply. That this theory does not provide an adequate theory of QSS in the SGS model 
is clear from the earliest studies of this issue~\cite{hohl+campbell_1968, goldstein+cuperman+lecar_1969,   lecar+cohen_1971}
which indeed used this model to probe the possible validity of LB theory for  3D gravitating systems. 
More recently a study of these questions in the SGS model has been reported by Yamaguchi \cite{yamaguchi_2008},
who finds reasonable agreement with LB in a certain range of initial conditions, and, like in the
work of Levin et al. mentioned above, proposes a modification of it to account for the QSS observed in 
other cases. We will compare in detail our results to these previous works.

Studies of the SGS model in the astrophysical  context go back at least  as far as that of \cite{camm}, 
and  there have been numerous studies of it also in the statistical mechanics literature in the decades since.
Many of these studies focussed on the question of relaxation to the thermal equiiibrium of the model,
for which the exact expression was first derived by Rybicki \cite{rybicki_1971}. 
That this relaxation, like in other long-range systems, takes place very slowly, on a timescale which 
diverges with the number of particles,  has been clear since the earliest works, but the precise time 
scale and parametric dependences thereof have been the subject of considerable study and even
some controversy (see e.g. \cite{tsuchiya+gouda+konishi_1996, miller_1996, yawn+miller_2003, mjtw_jstat_2010} and 
references therein). In a recent  work \cite{mjtw_jstat_2010} on this  question, we have established clearly that 
the relaxation time  from a range of initial conditions depends {\it linearly} on the number of particles $N$
\footnote{This result is consistent, notably, with the analysis of \cite{chavanis_kEqns_2010} 
which argues that a timescale linear in $N$ arises because of ``resonances''  
present in spatially inhomogeneous QSS in 1D systems, but not in spatially homogeneous 
QSS which occur in 1D systems such as the HMF model, where a faster scaling with $N$ is 
indeed observed (see, e.g., \cite{yamaguchi+barre+bouchet+dauxois+ruffo_2004}).}, while also 
showing a strong dependence on the intermediate QSS state (or states).
Besides the early and
more recent studies cited above which consider the QSS attained on the shorter mean-field time scales 
(i.e. through violent relaxation) and  LB theory, there are also studies \cite{mineau+rouet+feix_1990, rouet+feix} 
which argue that the assumption that QSS always result from mean-field dynamics
may not be always correct:  starting from certain initial conditions
the initial phase of relaxation is observed to lead to phase space densities which have
large holes which rotate in phase space, which persist on the time scales of the simulations.
In our analysis below we will examine this question carefully, as it is clearly of central
importance to understand whether the formation of a QSS is indeed a good description
of the outcome of violent relaxation if one is comparing with a theory which, by construction,
assumes such a outcome. 

The article is organized as follows. First we will start in section \ref{sim} with the
definition of the model, and its numerical integration. In the following 
section \ref{theory}  we review the theory of violent relaxation of Lynden-Bell 
and describe our calculations of the predictions for the density profiles, and
velocity and energy distributions . We will also introduce a simple set of
``order parameters"  which we use to characterize the QSS. In section \ref{ic}  we 
describe the specific class of initial conditions which we investigate.  
In section \ref{nume} we report our numerical results, comparing them
to the theoretical LB predictions. In the following section we confront
our results with two proposals which have been made in the recent 
literature to explain the properties of QSS when the LB is clearly
inadequate. We also discuss our results briefly in the light of 
the kinetic theory for collisionless relaxation developed in 
\cite{chavanis_KTql_2008} (and references therein). 
In our conclusions we summarize our findings and
conclusions, and  suggest some directions for further investigation
in both 1D and 3D self-gravitating systems.

\section{The self-gravitating sheet model} \label{sim}
We consider identical particles of mass $m$ in one dimension which are mutually attracted 
by a force independent of their separation, i.e.,  the force on a particle $i$ due to a particle $j$ 
is
\begin{equation}
F_{ij}=- gm^{2}\frac{x_{i}-x_{j}}{|x_{i}-x_{j}|} \equiv  - gm^2 \rm{sgn} ( x_{i}-x_{j}) \, \nonumber 
\end{equation} 
where $g$ is a couping constant. 
If the particles in one dimension are considered as infinitely thin parallel sheets in three
dimensions interating by 3D Newtonian gravity, it is simple to show that  $gm\equiv 2\pi \Sigma G $ where 
$G$ is Newton's constant and $\Sigma$ is the mass per unit surface area of the sheets. 
In a system of a finite number of such particles the total force acting on the $i^{th}$ particle
at any time may be expressed as 
\begin{equation}
F_{i}=gm^{2}[N^{i}_{+}-N^{i}_{-}] \label{total-force}
\end{equation} 
where $N^{i}_{+}$ denotes the number of particles on the right of $i^{th}$ 
particle and $N^{i}_{-}$ for the left. 

The fact that the force is thus constant other than when particles cross leads to
one of the very nice features of this toy model: its numerical integration requires
only the solution of algebraic (quadratic) equations to determine the time of 
the next particle crossing. This means that the only limit on the precision of
integration is that of the machine in solving such equations, and that no 
numerical parameters need be introduced. Another simplification
comes from the fact that, in one dimension, the crossing of two particles
without discontinuity in the velocities is, up to labelling of the particles,
equivalent to an elastic collision in which particles exchange velocities.
If we are not interested in following the trajectories of individual particles,
we can thus consider the system as consisting of particles on which
the forces are constant in time [and given by the initial value of  (\ref{total-force})], and which
undergo elastic collisions when they collide.
The optimal way to treat this 
kind of problem is, as has been pointed out and discussed in detail in \cite{noullez_etal},
by using a so-called ``heap-based'' algorithm, which uses an object
called a ``heap'' to store in an ordered way the next crossing 
times of the pairs. This algorithm requires a number of operations of 
order $\log(N)$ to  determine which of the $N-1$ pairs crosses
next. Given that the number of 
crossings per particle per unit time grows in proportion to $N$, the 
simulation time thus grows in proportion to $N^{2} \log(N)$. 
As is common practice we will use the
total energy (which is conserved in the continuum model) as a control parameter. For the
longest simulations we report the error in total energy of the order of $10^{-9}\%$.

\section{Predictions of Lynden-Bell theory} 
\label{theory}
In this section we very briefly recall the basics of the theory of Lynden-Bell, and
describe how we calculate its predictions for different quantities, in the case 
of waterbag initial conditions. 

A statistical theory to describe the stationary states arising from violent relaxation
through mean-field forces has  been proposed by Lynden-Bell in 1967~\cite{lynden_bell_1967}. 
Such states were proposed to arise from the relaxation of the coarse-grained phase space density 
to that derived by maximizing the entropy derived for the latter by ``counting" the 
(fine-grained or ``microscopic" ) phase space configurations states consistent with the conservation 
laws imposed by the collisionless (Vlasov)  
dynamics.  For the case of an initial ``water-bag" phase space density, i.e., in which
the microscopic phase space density has the same value everywhere it is non-zero, 
these conservation laws simply require the conservation of the phase space volume  
``occupied"  by this  constant density, $f_0$ say. The calculation of the entropy is then equivalent
to that for identical particles with a ``fermionic" exclusion,  and gives (in one dimension)
\begin{equation}
S [\bar{n}] =\int\int \bigg\{ \bar{n} \, \textrm{ln} \, \bar{n}+(1-\bar{n})\textrm{ln}(1-\bar{n}) \bigg\} dxdv \label{lbentropy}
\end{equation}
where $\bar{n}\equiv \bar{f}/f_0$, and  $\bar{f}$ is the coarse-grained phase space
distribution in the macrocell at $(x,v)$. 
Maximization of (\ref{lbentropy}) gives 
\begin{equation}
\bar{f}(x,v)=\frac{f_{0}}{1+e^{\beta (\epsilon (x,v)-\mu)}}  
\label{lbequation}
\end{equation}
where $\epsilon (x,v)=\frac{v^{2}}{2}+\varphi(x)$ denotes the energy density of phase-space element at $(x,v)$.
The constants $\beta$ and $\mu$ are Lagrange multipliers associated with the conservation of the total
mass $M$, and total energy $E$, of the system:
\begin{eqnarray}
M &=& \int\int \bar{f} (x,v) dxdv \label{mconstrain} \\
E &=& \int\int (\frac{v^{2}}{2}+\frac{\varphi}{2}) \bar{f} (x,v)dxdv\,, \label{econstrain}
\end{eqnarray}
where $\varphi(x)$ is the mean field potential generated by the mass density $\rho(x)$.
Except in the degenerate and non-degenerate limits, corresponding to $\beta \rightarrow \infty$ and 
$\beta \rightarrow 0$ respectively,  it is not possible to solve these equations analytically to
derive ($\beta$,$\mu$) for any given $M$, $E$ and $f_0$. It is, however, straightforward 
to do so numerically, as described in Appendix  \ref{findmubeta} (see also \cite{lecar+cohen_1971}).

We note that, although the prediction of LB theory for a water-bag initial condition 
depends in general on the three parameters $M$, $E$ and $f_0$,  for the SGS
model there is only one additional dimensionful quantity relevant in the
continuum limit, the coupling $g$.  Thus units can always be chosen so that two of 
$M$, $E$ and $f_0$ are fixed, and the LB prediction can therefore depend
non-trivially (up to a rescaling) only on {\it one parameter}.  A convenient choice of this
parameter, which we will use here, is
\begin{equation}
\xi_D \equiv \frac{E-E_D}{E_D}
\label{def-xiD}
\end{equation}
where $E_D (M, f_0)$ is the energy of the system with mass $M$ and phase
space density $f_0$ in the degenerate limit, i.e., $\xi_D$ is the energy of the system above the degenerate limit 
normalized to the lowest energy possible for the same mass 
and phase space density. The expression for $E_D$ is given in 
Appendix \ref{degenerate} (see also  \cite{lecar+cohen_1971}).

We next describe how we derive, once $\beta$ and $\mu$ are known, the
LB predictions for the various quantities we will measure in our simulations.

\subsection{Spatial distribution} \label{densiprofile}
Using (\ref{lbequation}), the Poisson equation gives, 
\begin{equation}
\frac{\partial^{2} \varphi (x)}{\partial x} =2g \rho(x) \equiv
2g \int_{-\infty}^{\infty}\frac{f_{0}}{1+e^{\beta(\frac{v^{2}}{2}+\varphi (x)-\mu)}} dv 
\label{eqdensiprofile_poisson}
\end{equation}
where $\rho(x)$ is the mass density profile (which we will refer to simply as 
the ``density profile").  It is simple numerically to solve this (second order differential) equation 
for $\varphi(x)$, and then to determine the mass density profile, using the 
boundary conditions $\frac{d \varphi}{d x}\large|_{x=0}=0$ and $\varphi\large|_{x=0}=0$.

\subsection{Velocity distribution} \label{veloprofile}
The velocity distribution may be written
\begin{equation}
\theta(v)=2\int_{0}^{\infty} \frac{f_{0}}{1+e^{\beta(\frac{v^{2}}{2}+\varphi-\mu)}}\cdot\frac{1}{a(\varphi)} d\varphi  
\label{eqveloprofile1}
\end{equation}
where 
\begin{equation}
a(x)=\frac{\partial \varphi (x)}{\partial x} \label{a1}
\end{equation}
is, up to a sign, the gravitational acceleration. 
Using the Poisson equation we have
\begin{equation}
\frac{d^{2} \varphi (x)}{d x^2}=\frac{1}{2}\frac{\partial (a^{2}(\varphi))}{\partial\varphi}=2g \rho \label{a2}
\end{equation}
and therefore
\begin{equation}
a(\varphi)=\sqrt{4g\int^{\varphi}_{0} \rho(\varphi ') d\varphi '}. \label{a3}
\end{equation}
Using the previously determined $\rho(\varphi)$ we obtain $\theta(v)$ using (\ref{eqveloprofile1}).

\subsection{Energy distribution} \label{enerprofile}
The distribution of particle energies is defined by
\begin{equation}
{F}(\epsilon)=\int \delta(\epsilon-[\frac{v^{2}}{2}-\varphi(x)])\cdot \bar{f}(x,v) dxdv \label{fe1}
\end{equation}
with 
\begin{equation}
\int F(\epsilon) d\epsilon = 1. \label{fe2}
\end{equation}
Integrating we obtain
\begin{equation}
F(\epsilon)=D(\epsilon)\bar{f}(\epsilon) \label{conversion}
\end{equation}
where
\begin{equation}
D(\epsilon)=\int_{0}^{\epsilon} \frac{1}{a(\varphi)}\cdot \frac{2\sqrt{2}}{\sqrt{\epsilon-\varphi}}d\varphi 
\label{density-states}
\end{equation}
is the density of states at energy $\epsilon$.

While the results for $\rho(x)$ and $\theta (v)$ do not depend on the 
choice of the zero point of the potential, this latter result does.
It is straightforward numerically, to use, rather than $\varphi \large|_{x=0}=0$, 
\begin{equation}
\varphi_0\equiv \varphi\large|_{x=0}=g\int_{-\infty}^{\infty} |x|\rho (x) dx, \label{phi0}
\end{equation}
i.e., that corresponding to a pair potential strictly proportional to the 
separation between particles. Given that $a(x)$ defined in (\ref{a1}) is necessarily positive 
for all $x\neq 0$, this is the minimum value of the potential
(and of the energy particle energy). Adapting this definition the 
energy distribution is still given by (\ref{conversion}), but with
$D(\epsilon)=0$ for $\epsilon < \varphi_0$ and 
\begin{equation}
D(\epsilon)=\int_{\varphi_0}^{\epsilon} \frac{1}{a(\varphi)}\cdot \frac{2\sqrt{2}}{\sqrt{\epsilon-\varphi}}d\varphi. 
\label{density-states-shifted}
\end{equation}

\subsection{Order parameters}
In order to characterize and compare the macroscopic properties of QSS 
it is convenient to calculate specific moments of the phase space distribution (rather than 
to study always the full distribution).  As discussed in \cite{mjtw_jstat_2010} a particularly 
relevant choice can be normalized ``crossed moments"  which give a measure 
of the ``entanglement" of the distribution in space and velocity coordinates,
by considering
\begin{equation}
\phi_{\alpha \beta} = \frac{\langle|x|^{\alpha}|v|^{\beta}\rangle}{\langle|x|^{\alpha}\rangle \langle |v|^{\beta}\rangle}-1 
\label{phi}
\end{equation}
for non-zero $\alpha$ and $\beta$, where
\begin{equation}
\langle u \rangle \equiv \frac{\int\int uf(x,v)dxdv}{\int\int f(x,v)dxdv}
\label{phi_def_lb}
\end{equation}
estimated in the discrete system with $N$ particles as
\begin{equation} 
\langle u \rangle \equiv \frac{1}{N} \sum_{i=1}^{N} u_{i}
\label{phi_def}
\end{equation}
where $u_i$ is the value measured for the particle $i$.
In thermal equilibrium the distribution function is separable, and so $\phi_{\alpha \beta}=0$.
Further it can be shown easily \cite{mjtw_jstat_2010} that the thermal equilbrium solution
at any energy is the unique separable stationary state, i.e., all QSS are non-separable.
Thus generically we expect these moments to be non-zero in a QSS (although 
any finite number of them can in principle vanish without implying separability).

\begin{figure}[!h]
\begin{center}
\begin{tabular}{c}
\hspace{-7mm}
\includegraphics[width=9cm]{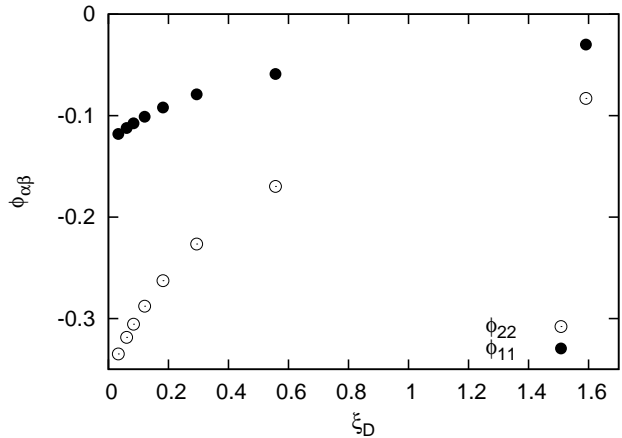}
\end{tabular}
\caption{The ``order parameters" $\phi_{11}$ and $\phi_{22}$ of the QSS
predicted by LB theory for waterbag initial conditions, plotted as a function 
of the normalized energy $\xi_D$.}
\label{phi_lb}
\end{center}
\end{figure}

Here we will use specifically the two moments  $\phi_{11}$ and  $\phi_{22}$ to
characterize and compare the QSS we obtain in our numerical simulations,
complemented when necessary by examination of the functions derived
above and in some cases of the full phase space density.
Given the LB solutions determined above (for waterbag initial conditions)
it is straightforward to calculate numerically the values of $\phi_{11}$ and 
$\phi_{22}$ predicted by LB for this case.
These are shown in Fig.~\ref{phi_lb} as a function of the parameter
$\xi_D$ (which, as discussed above, can be taken as the single 
parameter on which the LB result depends).  We note that both 
parameters are always negative but increase towards zero as
we go to the non-degenerate limit. Indeed in this limit the LB
prediction tends to the (separable) thermal equilibrium solution. 

\section{Initial conditions} \label{ic}

In our numerical study we consider particles distributed initially by randomly
sampling different classes of waterbag initial conditions, i.e., in which the 
phase space density takes the same value  $f_{0}$ everywhere it is 
non-zero.  Specifically we consider, in order:

\begin{figure}[t!]
\begin{center}
\begin{tabular}{cc}
\hspace{-6mm}
\includegraphics[width=4.5cm]{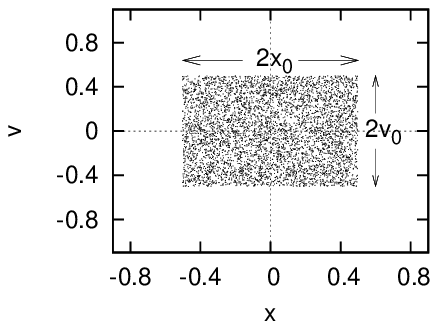} \hspace{-4mm} & 
\includegraphics[width=4.5cm]{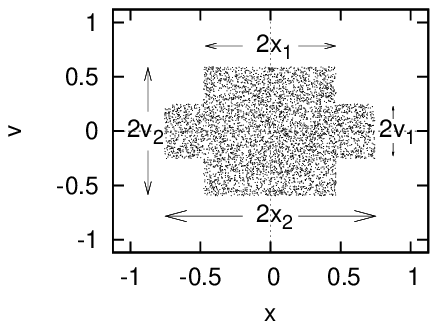}
\end{tabular}
\caption{Realizations with $N=5000$ particles of an SRW initial condition 
(left panel) and DRW initial condition (right panel). The two configurations 
have the same  value of $\xi_D$ (up to finite $N$ corrections). The units
used here are specified at the beginning of Sec.~\ref{nume} below.
}
\label{exbag}
\end{center}
\end{figure}

\begin{itemize}
\item {\bf Single rectangular waterbags (SRW)}, in which the support of the initial phase
space density is a rectangle centred on the origin, i.e.,
\begin{equation}
f(x,v) =  f_0 \Theta (x_0 -x)\Theta (x_0 +x)\Theta (v_0 -v) \Theta (v_0 +v)
\label{sr-waterbag}
\end{equation}
where $\Theta$ is the Heaviside function. As, in the continuum limit, the only 
parameters in the problem are then four ---
$f_0$, $x_0$, $v_0$ and the coupling $g$ --- there is in fact only
one relevant parameter characterizing the system once units are chosen.
A natural physical choice of this parameter is the initial virial ratio $R_0$,
which a simple calculation shows is given by
\begin{equation}
R_{0}\equiv \frac{2T_{0}}{U_{0}}=\frac{v_{0}^{2}}{gMx_{0}} \label{virialratio}
\end{equation}
where $T_0$ and $U_0$ are the initial kinetic and potential energies given by
\begin{equation}
T_{0} = \frac{1}{6}Mv_{0}^{2} \ \ \ , \ \ \ U_{0} = \frac{1}{3}gM^{2}x_{0}. \label{t_u}
\end{equation}
An example of such a configuration with $R_{0}=0.5$ is given in the left panel 
of Fig.~\ref{exbag}. 

As discussed above the LB prediction also depends on only one 
parameter, which we can take to be $\xi_{D}$, the ratio of the energy of
the configuration to that of the degenerate limit of LB (i.e. the minimum 
allowed energy of the given mass at phase space density $f_0$).
The energy and mass in the limit of a degenerate system are given 
as functions of $\mu$ by (\ref{degenerate13})  and (\ref{degenerate5}). 
Eliminating $\mu$ we obtain
\begin{equation}
E_{D} = \frac{B(\frac{3}{2},\frac{2}{3})}{12^{\frac{1}{3}}}gx_{0}M^{2}R_{0}^{\frac{1}{3}} \label{ed_md}\,, 
\end{equation}
and thus 
\begin{equation}
\xi_{D} = \frac{E}{E_{D}} = \frac{12^{\frac{1}{3}}}{3B(\frac{3}{2},\frac{2}{3})}
(\frac{1}{R_{0}^{\frac{1}{3}}}+\frac{R_{0}^{\frac{2}{3}}}{2}) = 0.688(\frac{1}{R_{0}^{\frac{1}{3}}}
+\frac{R_{0}^{\frac{2}{3}}}{2})\,
\label{xi_r0_SRW}
\end{equation}
where $B(\frac{3}{2},\frac{2}{3})$ is a beta function. This expression is 
is plotted in Fig.~\ref{fig_xid_r0}. The SRW with $R_{0}=1$ 
is thus the lowest energy configuration, and there are otherwise
two values of $R_0$ for each  value of $\xi_D$.

\begin{figure}[!h]
\begin{center}
\begin{tabular}{c}
\includegraphics[width=8cm]{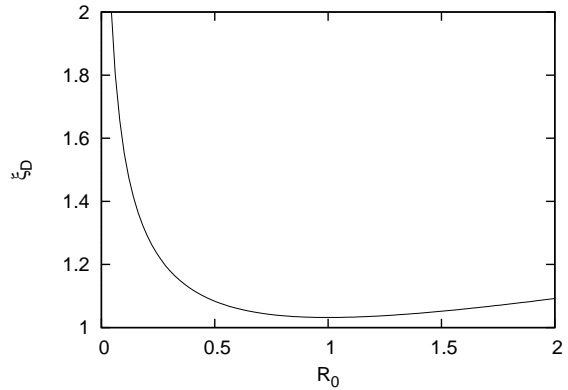}
\end{tabular}
\caption{$\xi_{D}$ as a function of $R_{0}$ for a SRW initial condition.} 
\label{fig_xid_r0}
\end{center}
\end{figure} 

\item{\bf Double rectangular waterbag (DRW)},  
\label{doublewb}
in which the support of the initial continuum phase space density 
is like that shown in the right panel of Fig.~\ref{exbag}: 
\begin{eqnarray}
f(x,v) &=&  f_0 \Theta (x+ x_1) \Theta (x_1 -x) \Theta (v+v_1)  \Theta (v_1 -v) \nonumber \\
&+& f_0 \Theta (x+x_2) \Theta (-x_1 -x) \Theta (v+v_2)   \Theta (v_2 -v) \nonumber\\
&+& f_0 \Theta (x-x_2)  \Theta (x_1-x)   \Theta (v+v_2)  \Theta (v_2 -v) \,.\nonumber
\end{eqnarray}
As this has two additional parameters compared to the SRW, it is 
effectively a {\it three} parameter family of initial conditions, which 
coincides with the SRW when $v_1=v_2$, $x_1=0$ or $x_1=x_2$.
When they differ from
the SRW, they are spatially inhomogeneous, with a ratio
of densities $\delta=\frac{v_{1}}{v_{2}}$ in the two different regions.
We will choose to characterize them by this parameter, together
with $\xi_D$ and the initial virial ratio $R_0$. LB theory 
thus predicts that the final state should be independent of 
$R_0$ and $\delta$ at given $\xi_D$. The relevant expressions
for the kinetic and potential energies of the DRW configuration
are given in  Appendix \ref{appen_db}.

\item{\bf Disjoint waterbags (DW)}, in which the initial phase space density
is made of two disjoint regions with simple shapes, either rectangular
or elliptical.  We will use such configurations to further explore some
of the conclusions draw from the study of the SRW and DRW 
configurations.

\end{itemize} 

\section{Numerical results} 
\label{nume}

\subsection{Choice of units} 
\label{units}

Unless otherwise indicated our results will be given
in units fixed by taking $g=1$,  $M=1$,
and $L_0=1$ where $L_0$ the 
{\it initial linear size of the system}, i.e.,  the distance between 
the outer extremities of the theoretical waterbag initial 
condition. This implies that the unit of time is
\begin{equation}
t_{c}=\frac{1}{\sqrt{g\rho_{0}}}  
\label{tscale}
\end{equation}
where $\rho_0=M/L_0$ is the initial mean mass density. 
This is simply a characteristic time scale  for a particle to 
cross the system.  In the cold limit (i.e. with zero initial 
velocity, with $R_0 \rightarrow 0$) of the SRW initial 
conditions, it corresponds exactly to the time in which all 
the mass falls to the centre of the  system.

\subsection{Attainment of QSS and their characterization: generalities}

That the SGS model with a large number of particles --- 
just as such 3D self-gravitating and  other long range 
interacting systems which have been studied in the literature ---  
give rise typically to QSS starting from initial conditions such as 
those above has been discussed elsewhere in many studies (see
references given in the introduction). The attainment of a QSS should be tested, in 
theory, by considering 
the full phase space density coarse-grained at some chosen
scale. One would then verify whether its evolution after some
initial period (of violent relaxation) tends to 
\begin{equation}
\bar{f} (x,v,t) = \bar{f}_{QSS} (x,v) + \delta \bar{ f} (x,v,t) 
\label{f-qss}
\end{equation}
where the amplitude of the fluctuations $|\delta \bar{ f} (x,v,t)|$
decreases as $N$ increases. For our study here, in which we consider 
how the properties of  these QSS depend on the initial conditions, what
is of importance is that we evolve the corresponding system to a time at
which the approximation  (\ref{f-qss}) indeed holds well, for $N$ sufficiently
large so that the fluctuations  $\delta \bar{ f} (x,v,t)$ introduce a
negligible uncertainty into the quantities used to characterize
the QSS. 

In practice numerical limitations on $N$ make a direct 
analysis extremely difficult, and one typically considers the behavior 
of single macroscopic  parameters, such as the virial ratio, or the 
magnetization in models (e.g. the HMF model) where it is defined.
This is then complemented by a visual inspection of the system 
represented in phase space.  To describe the properties of the 
QSS one then considers typically the density profiles, velocity and/or 
energy distribution. We have shown in \cite{mjtw_jstat_2010},
where we studied the very long time behaviour of QSS resulting 
from SRW initial conditions,  that the parameters $\phi_{11}$ and 
$\phi_{22}$ defined  above are very useful macroscopic ``order 
parameters", which can be used to diagnose both the attainment 
of a QSS and to characterize this state.  We will use them
here for the same purpose, supplementing their calculation
where necessary, or interesting,  by a fuller analysis of the
distribution functions.  

\begin{figure}[t!]
\begin{center}
\begin{tabular}{c}
\includegraphics[width=7cm]{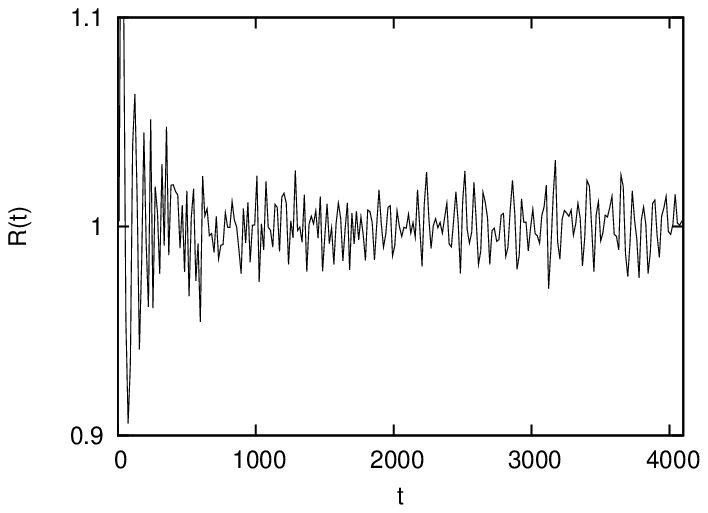} \\
\includegraphics[width=7cm]{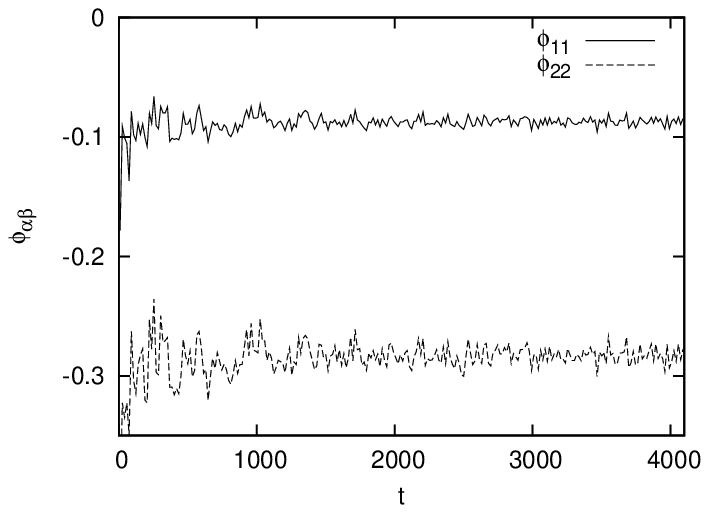} 
\end{tabular}
\caption{Temporal evolution of virial ratio (top panel), and $\phi_{11}$ and $\phi_{22}$
(lower panel) starting from a realization with $N=10^4$ particles of a DW 
initial condition (shown in first panel of Fig. \ref{phase_space_hole}). The time
units here are such that $t_c=\sqrt{3/2}$, i.e., $t=10 \approx 8.2 t_c $.
} \label{phi_case01}
\end{center}
\end{figure}

\begin{figure}[t!]
\begin{center}
\begin{tabular}{cc}
\hspace{-6mm}
\includegraphics[width=4.7cm]{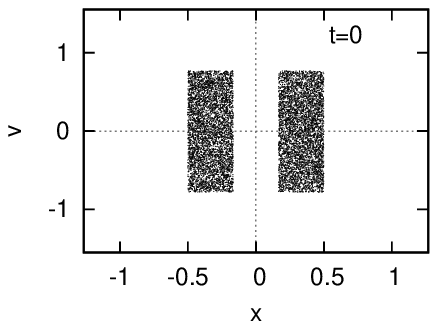} \hspace{-4mm} & 
\includegraphics[width=4.7cm]{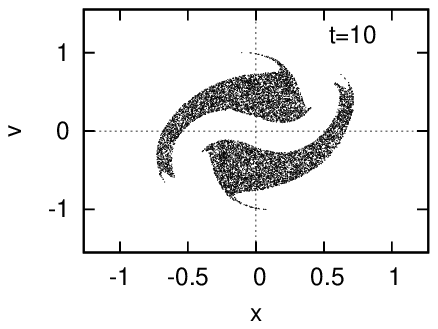} \\
\hspace{-6mm}
\includegraphics[width=4.7cm]{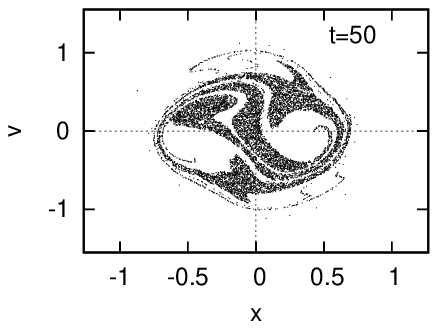} \hspace{-4mm} & 
\includegraphics[width=4.7cm]{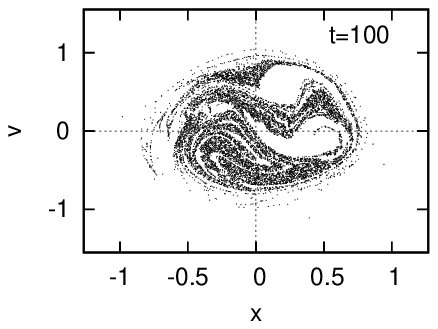} \\
\hspace{-6mm}
\includegraphics[width=4.7cm]{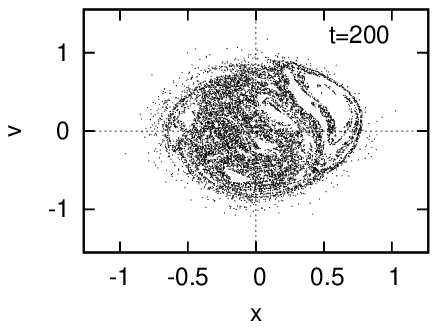} \hspace{-4mm} & 
\includegraphics[width=4.7cm]{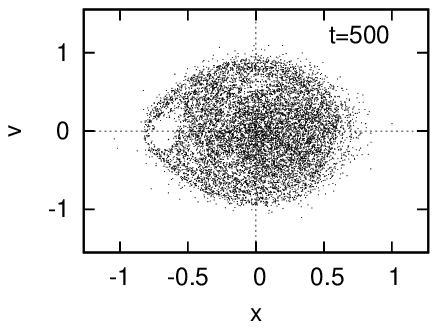} \\
\hspace{-6mm}
\includegraphics[width=4.7cm]{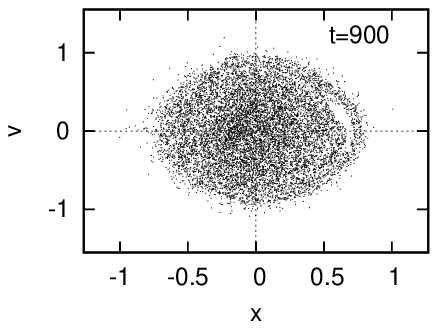} \hspace{-4mm} & 
\includegraphics[width=4.7cm]{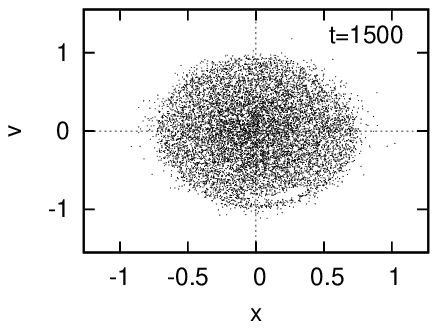} \\
\hspace{-6mm}
 \includegraphics[width=4.7cm]{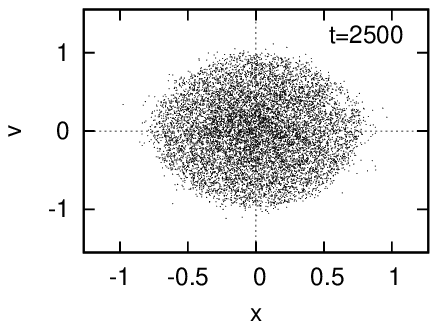} \hspace{-4mm} & 
 \includegraphics[width=4.7cm]{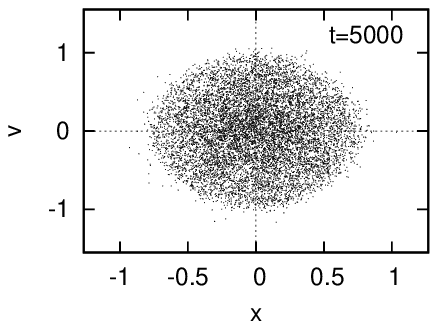} 
\end{tabular}
\caption{Phase space plot of particle trajectories evolved from the DW 
initial condition shown in the first panel (with $N=10^4$ particles).
The time units are those indicated in the previous figure.} 
\label{phase_space_hole}
\end{center}
\end{figure}

To determine whether a QSS is reached, and on what time
scale, we thus study firstly the evolution of the virial ratio,
and of $\phi_{11}$ and $\phi_{22}$.  While the characteristic
time for the mean-field dynamics is of order $t_c$ defined 
above, the completion of relaxation to QSS (in the sense
defined above) takes typically of order several tens to
one hundred $t_c$ for SRW initial conditions.
Further this time depends, unsurprisingly, on 
the nature of the initial condition, with very cold initial
conditions --- further from virial equilibrium initially ---
taking significantly longer to relax.

For DRW and DW initial conditions we observe even
greater variation in the time for full relaxation to a QSS 
than for SRW, with, in some cases, significant 
persistent fluctuations in the macroscopic parameters.  
An example of such a case is shown in 
Fig.~\ref{phi_case01}, in which the upper panel
shows the evolution of the virial ratio and the
lower panel that of the parameters 
$\phi_{11}$ and $\phi_{22}$, for a DW initial
condition sampled with $N=10^4$ particles.
The full phase space plot is shown in 
Fig.~\ref{phase_space_hole}.
This reveals that it is a persistent  ``rotating hole" 
feature in the phase space which gives rise to
the (small but clearly visible) coherent fluctuations
in the averaged parameters in Fig.~\ref{phi_case01}.
This is precisely the kind of effect which has
been documented in the two studies 
\cite{mineau+rouet+feix_1990, rouet+feix} 
mentioned in the introduction, and which
has been argued in this context to show 
that LB theory is incorrect (as it predicts,
by construction, the attainment of a
time independent phase space density).
While the hole we observe is clearly visible 
at  $t=500$ and indeed rotates in phase space, 
the subsequent two panels show that it slowly
disappears on a time scale of order a few thousand 
dynamical times. Thus it appears that 
the relaxation of these holes simply represents
a prolongation of the {\it collisionless} relaxation to 
a well defined QSS,  as no tendency of the
system to evolve towards thermal equilibrium 
(corresponding to $\phi_{11}$ and $\phi_{22}$
equal to zero)  is evidenced on this time scale. 
Further study, however, would be required to 
establish this conclusion more definitively for a broader
range of initial conditions, and to exclude
notably that collisional relaxation may play
some role. 

\subsection{SRW initial conditions}

\label{numesingle}
\begin{figure}[!h]
\begin{center}
\begin{tabular}{cc}
\hspace{-5mm}
\includegraphics[width=4.7cm]{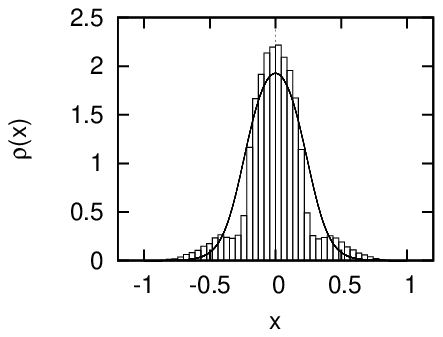}   \hspace{-5mm}
& \includegraphics[width=4.7cm]{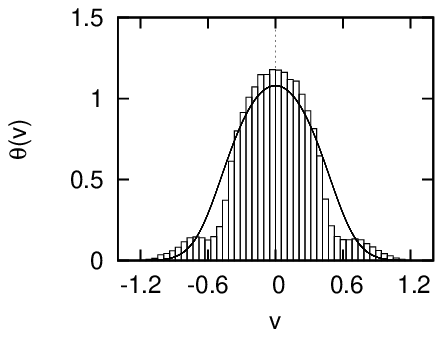} 
\end{tabular}
\includegraphics[width=5.5cm]{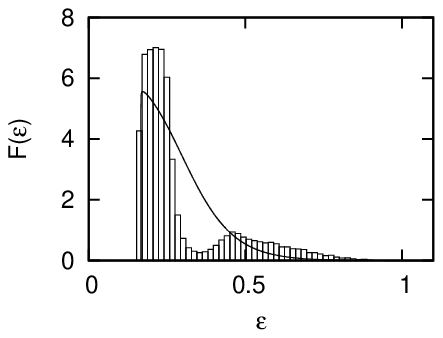}
\caption{
The density profile (top left), velocity distribution (top right) and energy distribution (bottom) 
for the QSS obtained starting from SRW with $R_{0}=0.1$.  The solid lines are the
corresponding LB predictions.} \label{single01}
\end{center}
\end{figure}  

\begin{figure}[!h]
\begin{center}
\hspace{-5mm}
\begin{tabular}{cc}
\hspace{-5mm}
\includegraphics[width=4.7cm]{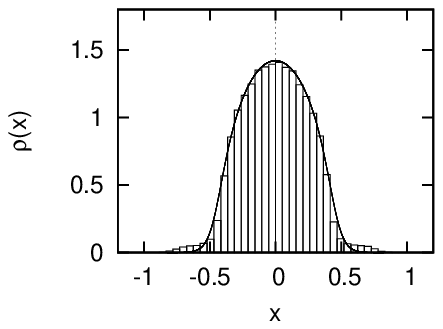} \hspace{-5mm}
& \includegraphics[width=4.7cm]{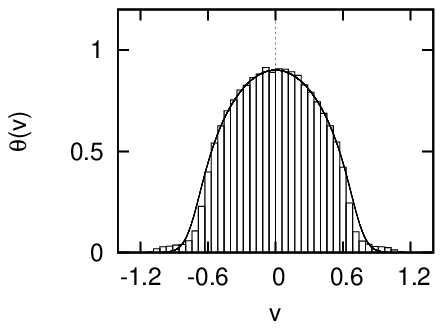} 
\end{tabular}
\includegraphics[width=5.5cm]{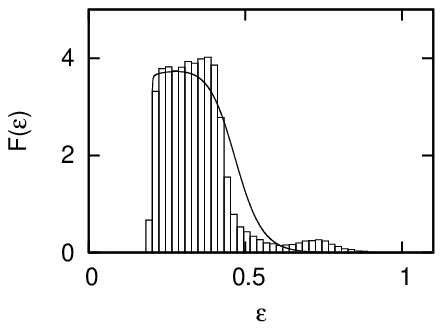}
\caption{The density profile(top left), velocity distribution (top right) and energy distribution (bottom) 
for the QSS obtained starting from SRW with $R_{0}=0.5$.  The solid lines are the
corresponding LB predictions.} \label{single05}
\end{center}
\end{figure}

\begin{figure}[!h]
\begin{center}
\begin{tabular}{cc}
\hspace{-5mm}
\includegraphics[width=4.7cm]{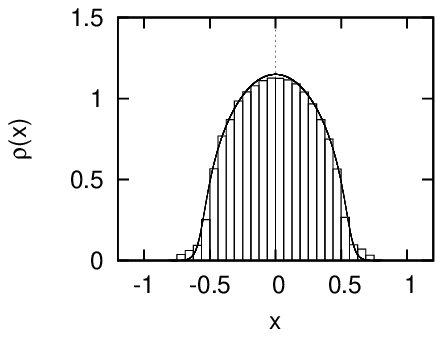}   \hspace{-5mm}
& \includegraphics[width=4.7cm]{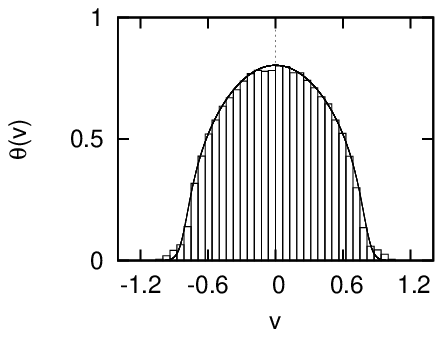} 
\end{tabular}
\includegraphics[width=5.5cm]{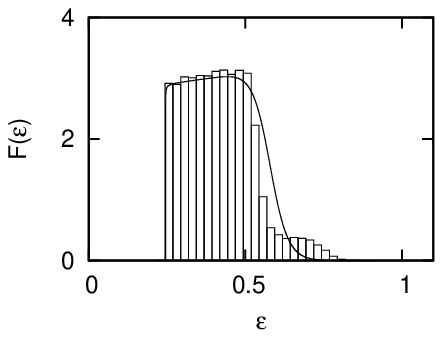}

\caption{
The density profile(top left), velocity distribution (top right) and energy distribution (bottom) 
for the QSS obtained starting from SRW with $R_{0}=1$.  The solid lines are the
corresponding LB predictions.} \label{single10}
\end{center}
\end{figure}

The density profiles, velocity distributions and energy distributions in the QSS obtained 
starting from SRW configurations with $R_{0}=0.1, 0.5, 1$ are shown in 
Figs.~ \ref{single01}, \ref{single05} and \ref{single10}.  These correspond to
averages over $30$ realizations of each initial condition sampled
with $N=5000$ particles, taken at $t=200 t_c$, by which time the QSS is
well established. 
In each case the LB predictions given in
Sec.~ \ref{theory} are shown also, corresponding to $\xi_{D}=0.56,0.08, 0.03$ 
respectively.  As observed already in early studies 
\cite{goldstein+cuperman+lecar_1969, lecar+cohen_1971} the qualitatively
most striking deviation from the prediction of LB theory is marked by
the appearance  of a ``core-halo" structure, most clearly evident in
the energy distribution obtained from the $R_0=0.1$ initial condition.
On the other hand, as underlined in the more recent study of 
\cite{yamaguchi_2008} for these same initial conditions, the agreement
of the LB theory with the observed QSS is in fact quite good for the case
$R_0=1$, while the discrepancy progressively increases as $R_0$ 
deviates from unity and a core-halo type structure appears. 

\begin{figure}[!h]
\begin{center}
\begin{tabular}{c}
\includegraphics[width=7cm]{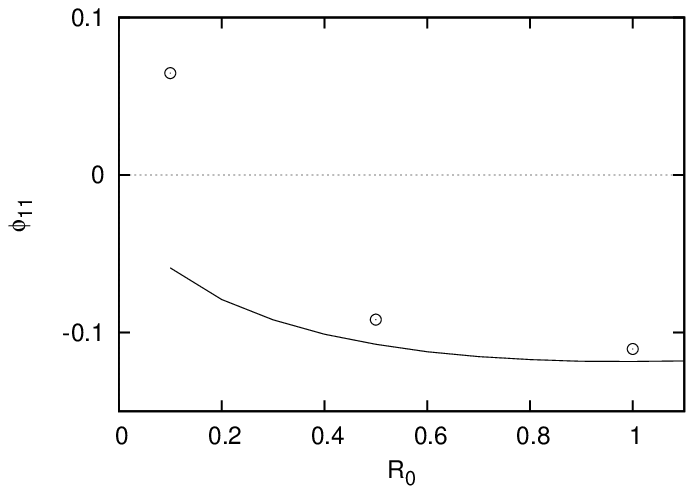} \\
\includegraphics[width=7cm]{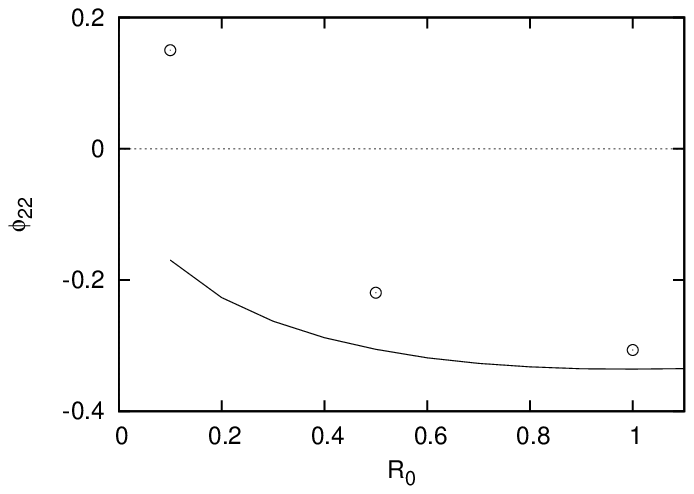} 
\end{tabular}
\caption{$\phi_{11}$(top) and $\phi_{22}$(bottom) in QSS as the function of $R_{0}$. The line indicates 
$\phi_{\alpha \beta}$ calculated by LB stationary state and the circle is the value obtained by numerical 
simulation.} \label{phi_lb_nume}
\end{center}
\end{figure}

Shown in  Fig.~\ref{phi_lb_nume} are the values of the parameter $\phi_{11}$ and $\phi_{22}$
in the QSS, and the values predicted by LB theory.  This plot summarizes in a simple
manner the conclusions above: the theory works quite well quantitatively  at the lowest 
energy state corresponding to $R_0=1$, but deviates greatly as we go towards the
less degenerate initial states.  Further the plot shows that the theory gives very
qualitatively the correct behavior of the parameters --- they increase monotonically 
with the initial $\xi_D$. At low degeneracy the sign of these parameters is a result
of the formation of a core which is colder than predicted: there is in this case an
excess of low velocity particles at small $x$.  

We note that these single parameters, $\phi_{11}$ and $\phi_{22}$,
actually allow a better diagnosis of the closeness to LB theory than the examination
of the full density or velocity distribution functions. Indeed comparing just these
two latter functions with the LB predictions,  we might conclude that the 
agreement is almost perfect. The energy distribution, on the other hand, allows
one to see clearly the discrepancies, which are then reflected well in
$\phi_{11}$ and $\phi_{22}$ \footnote{This "efficiency" of these parameters
as diagnostic tools was noted in \cite{mjtw_jstat_2010}, where it was 
shown, notably, that they could identify clearly stationary states arising
from certain initial conditions as QSS rather than the thermal equilibrium 
states which previous studies \cite{luwel+severne+rousseeuw_1983}
had mistakenly inferred them to be based on an analysis 
using $\rho(x)$ and $\theta(v)$.}  When considering a larger space of
initial conditions, as we do now, it is very convenient  to use these
parameters as diagnostics of the validity of LB.

\subsection{DRW initial conditions} 
\label{numedouble}

As described above the DRW initial conditions allow us to test further a basic prediction
of LB theory: the same QSS should result starting from any initial configuration in the 
range of  accessible ``microstates" at given mass and energy. For 1D gravity and
waterbag initial conditions, this means the QSS obtained should be the same
at a given $\xi_D$ independently of the shape of the waterbag.  As discussed
the DRW gives us a two dimensional space of such configurations, which
we choose to parametrize by the initial virial ratio  $R_0$ and density
contrast $\delta$.

\begin{figure}[!h]
\begin{center}
\begin{tabular}{c}
\includegraphics[width=7cm]{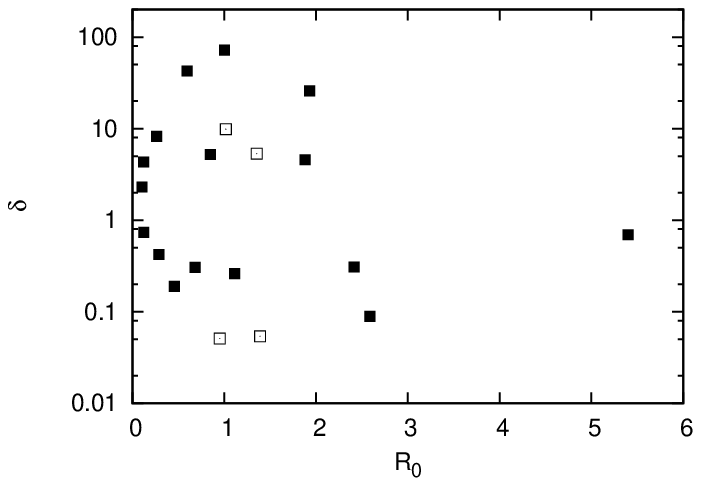} \\
\includegraphics[width=7cm]{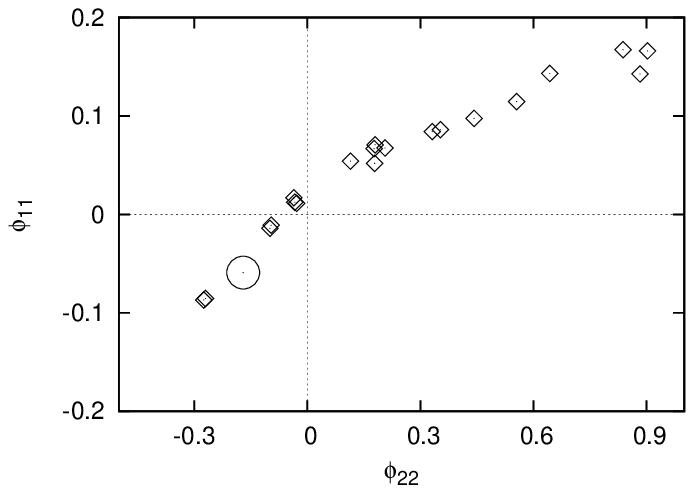}
\end{tabular}
\caption{The upper plot represents the twenty different DRW initial conditions
with $\xi_D=0.56$ (i.e. equal to that of SRW with $R_0=0.1$) according to their 
values of $R_0$ and $\delta$. The lower plot represents the values
of $(\phi_{22},\phi_{11})$ measured in the resulting QSS.  The
LB prediction lies at the centre of the small circle. The unfilled points
in the upper plot correspond to the four initial conditions which give
QSS closest to the LB prediction.}
\label{double01}
\end{center}
\end{figure}

\begin{figure}[!h]
\begin{center}
\begin{tabular}{c}
\includegraphics[width=7cm]{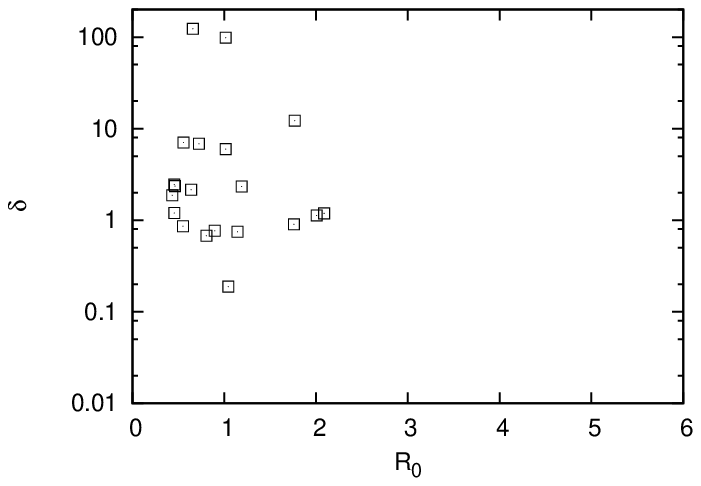} \\
\includegraphics[width=7cm]{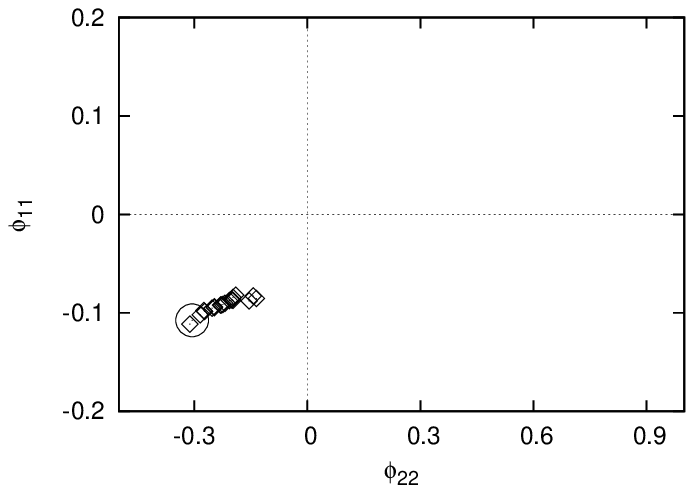} 
\end{tabular}
\caption{Same as in Fig.~\ref{double01}, but for DRW initial conditions
with $\xi_D=0.08$, i.e.,  equal to that of SRW with $R_0=0.5$. We keep
the scale as in the previous figure for easier comparison.}
\label{double05}
\end{center}
\end{figure}

\begin{figure}[!h]
\begin{center}
\begin{tabular}{c}
\includegraphics[width=7cm]{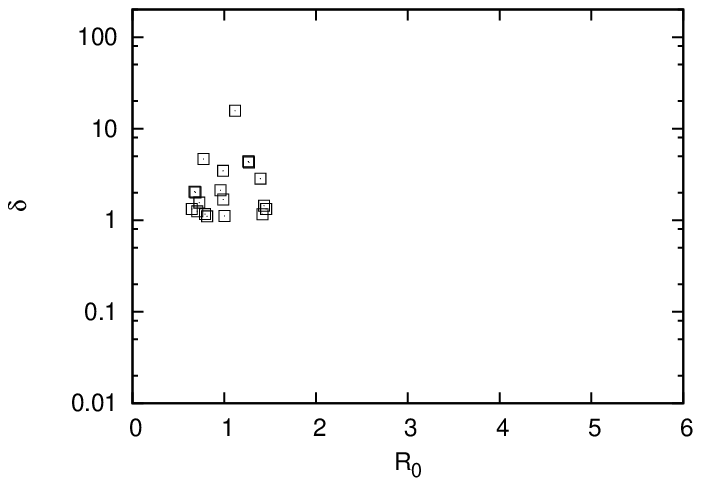} \\
\includegraphics[width=7cm]{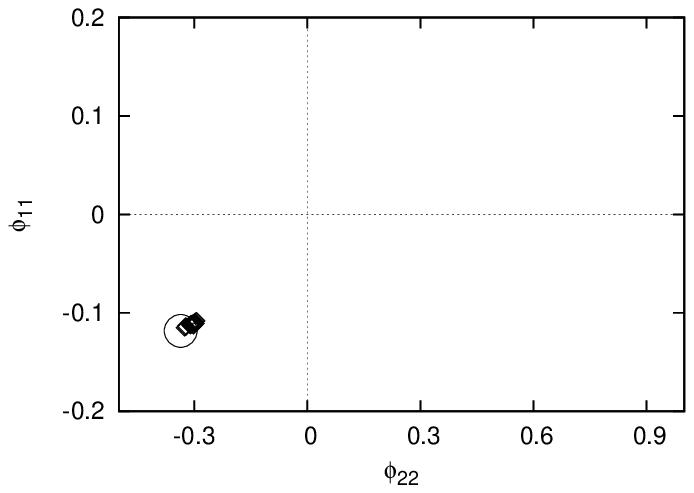} 
\end{tabular}
\caption{Same as in Fig.~\ref{double05}, but for DRW initial conditions
with $\xi_D=0.03$, i.e.,  equal to that of SRW with $R_0=1$. Same
scale as in Fig.~\ref{double01} for ease of comparison.} 
\label{double10}
\end{center}
\end{figure}

For each of the three values of $\xi_D$ corresponding to the SRW initial conditions 
above, we have simulated twenty different initial conditions chosen to explore
the available ($R_0$, $\delta$) space.  In each of 
Figs.~\ref{double01}, \ref{double05} and \ref{double10} are shown 
two plots: one shows the initial conditions in the  ($R_0$, $\delta$) plane
at the given value of $\xi_D$, the other the QSS obtained
from them as represented in the plane  ($\phi_{11}$, $\phi_{22}$).
The results are, as for the SRW above,  averages over $30$ realizations of 
each initial condition sampled with $N=5000$ particles, taken at 
$t \approx 200 t_c$. The fact that the spread in values of  ($R_0$, $\delta$) is much
smaller at smaller $\xi_D$ is simply a reflection of the fact that
as one goes towards the degenerate limit $\xi_D=0$ the constraints
limit the possible deformations more and more.

\begin{figure}[!h]
\begin{center}
\begin{tabular}{cc}
\hspace{-5mm}
\includegraphics[width=4.7cm]{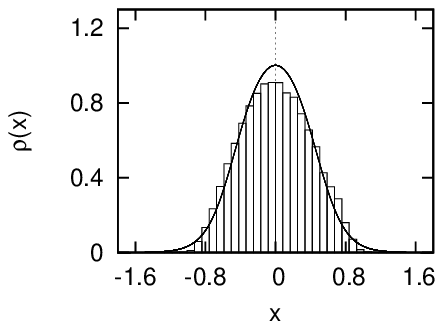}   \hspace{-5mm}
& \includegraphics[width=4.7cm]{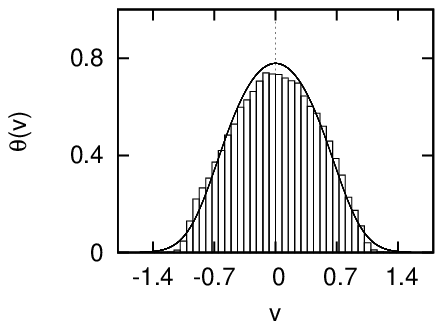} 
\end{tabular}
\includegraphics[width=5.5cm]{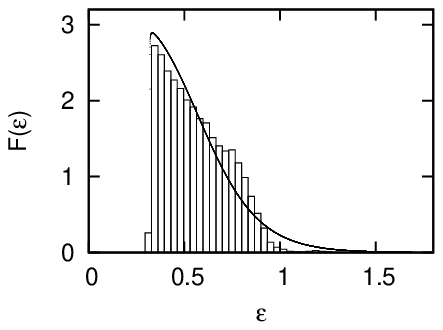} 
\caption{Density profile(top-left), velocity (top-right) and energy distribution (bottom) for 
DRW initial conditions with $\xi_D=0.56$ (i.e. the same energy as the SRW with 
$R_{0}=0.1$), $R_{0}=1.39$ and $\delta=0.054$.} \label{noncorehalo1}
\end{center}
\end{figure}

\begin{figure}[!h]
\begin{center}
\begin{tabular}{cc}
\hspace{-5mm}
\includegraphics[width=4.7cm]{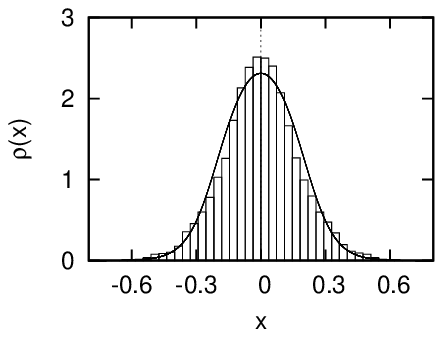}   \hspace{-5mm}
& \includegraphics[width=4.7cm]{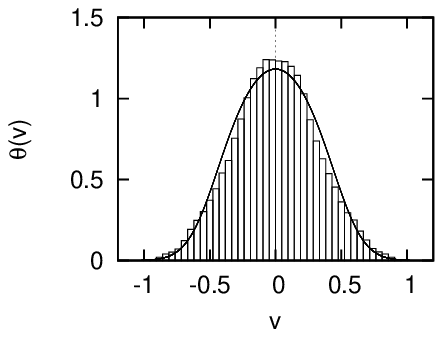} 
\end{tabular}
\includegraphics[width=5.5cm]{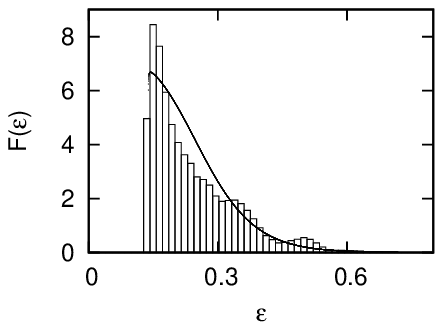} 
\caption{Density profile(top-left), velocity (top-right) and energy distribution (bottom) for 
DRW initial conditions with $\xi_D=0.56$ (i.e. the same energy as the SRW with 
$R_{0}=0.1$),  $R_{0}=1.017$ and $\delta=9.861$.} \label{noncorehalo2}
\end{center}
\end{figure}

In continuity with what we observed for the SRW, the results
show that LB theory works reasonably well at the two lower values
of $\xi_D$ --- the QSS varies only very little over the range
of different initial conditions --- it is grossly violated as we
go towards the non-degenerate limit.  Indeed the order 
parameters for QSS obtained starting from the same 
$\xi_D$ can differ in sign. Direct analysis of the distribution
functions confirms that this corresponds to QSS which are
completely different. On the other hand, certain initial
conditions at $\xi_D=0.56$ --- those corresponding to the
unfilled points in the upper plot of Fig.~\ref{double01} --- 
do appear to give QSS close to the LB prediction. To assess 
whether this is  really the case the density profiles, velocity and energy
distribution functions for two of these are shown in
Figs. ~\ref{noncorehalo1} and \ref{noncorehalo2}. 
While the agreement with the theoretical curves is
not perfect, it is comparable than that obtained for 
the initial conditions with $\xi_D=0.03$  --- indeed
the discrepancy between the LB prediction and the
observed distributions is no more than observed
above for the SRW initial conditions with 
$\xi_D=0.03$.

The strong deviations from the LB prediction, just as in the SRW,
manifest themselves in the shift towards positive values of
$\phi_{11}$ and $\phi_{22}$. Direct inspection of the 
distribution function of energy shows that this reflects again 
in all cases the appearance of a pronounced 
core-halo type structure.  Inspection of the plot of
the initial conditions in the  ($R_0$, $\delta$) space for
$\xi_D=0.56$ shows that all the cases which
approach LB (unfilled points) are characterized by an initial virial
ratio near unity, while the density contrast parameter $\delta$
appears to be irrelevant. On the other hand $R_0 \approx 1$
is clearly not a sufficient condition to guarantee agreement
with LB.


These results suggest therefore that LB theory works reasonably well always 
near the degenerate limit, and also for much higher energies for very 
specific initial conditions. In these cases, which seem to correlate
strongly with an initial virial ratio near unity,  the formation of a core-halo 
structure, not predicted by LB theory,  is avoided. 

\subsection{DW initial conditions}

\begin{figure}[!h]
\begin{center}
\begin{tabular}{cc}
\hspace{-6mm}
\includegraphics[width=4.7cm]{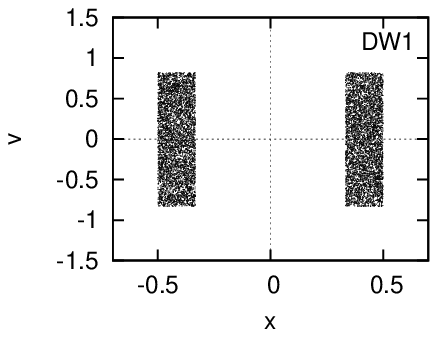} \hspace{-5mm} &
\includegraphics[width=4.7cm]{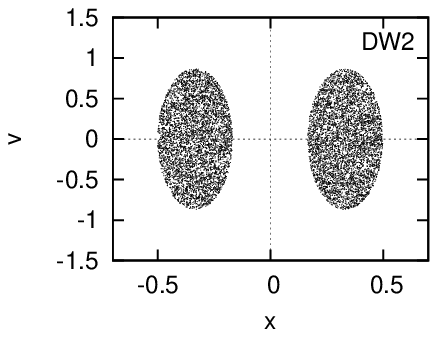} \\   
\hspace{-6mm}
\includegraphics[width=4.7cm]{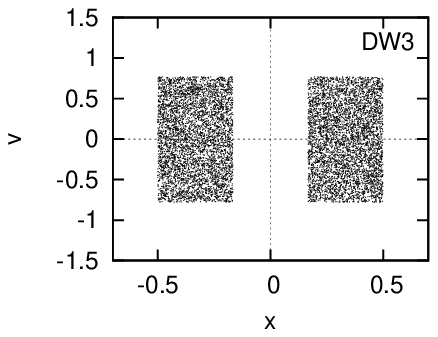} \hspace{-5mm} & 
\includegraphics[width=4.7cm]{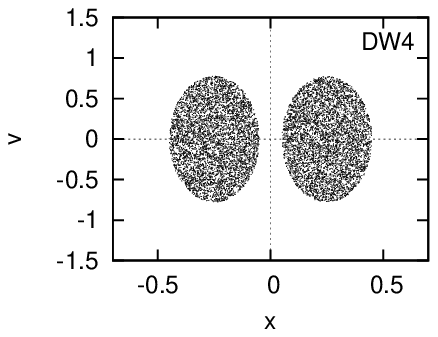} 
\end{tabular}
\caption{Four disjoint waterbag initial conditions with the number of case indicated in the panel. The 
corresponding $\xi_{D}$ are equal to $1.59, 0.58, 0.49$ and $0.23$ for case 1 to 4 respectively.}
\label{crossing_ic}
\end{center}
\end{figure}

\begin{figure}[!h]
\begin{center}
\includegraphics[width=8cm]{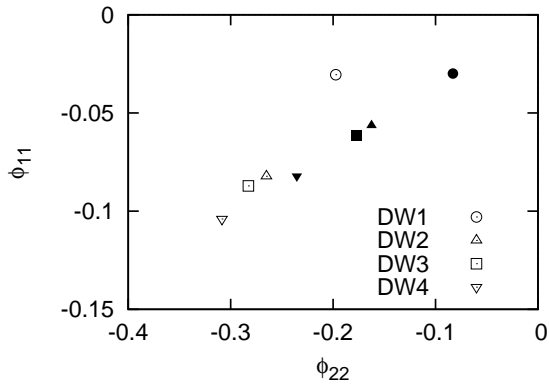}
\caption{ $\phi_{22}$ and $\phi_{11}$ of the QSS obtained from the initial conditions in
the previous figure. The unfilled symbols corresponds to the values obtained from 
numerical simulations, while the filled symbols are the LB predictions. }
\label{crossing_phi}
\end{center}
\end{figure}

\begin{figure}[!h]
\begin{center}
\begin{tabular}{cc}
\hspace{-5mm}
\includegraphics[width=4.7cm]{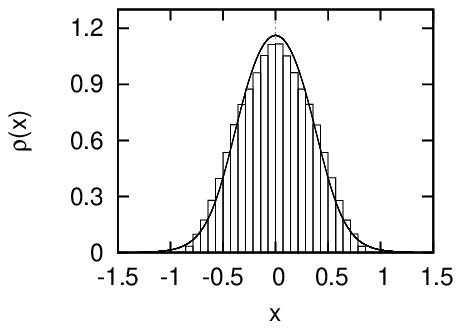}   \hspace{-5mm}
& \includegraphics[width=4.7cm]{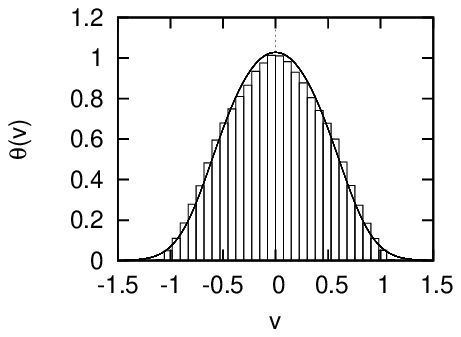} 
\end{tabular}
\includegraphics[width=5.5cm]{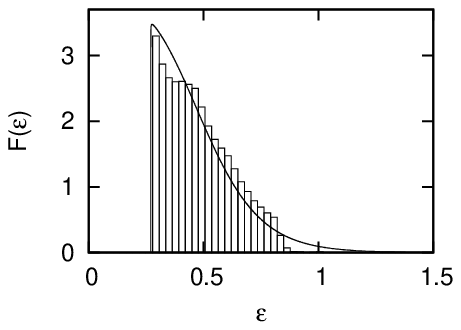}
\caption{The density profile (top left), velocity distribution (top right) and energy distribution (bottom) 
for the QSS obtained starting from the DW2 initial conditions ($\xi_{D}=0.58$). The solid curves
lines are the LB predictions.} \label{pdfs_DR3}
\end{center}
\end{figure}

\begin{figure}[!h]
\begin{center}
\begin{tabular}{cc}
\hspace{-5mm}
\includegraphics[width=4.7cm]{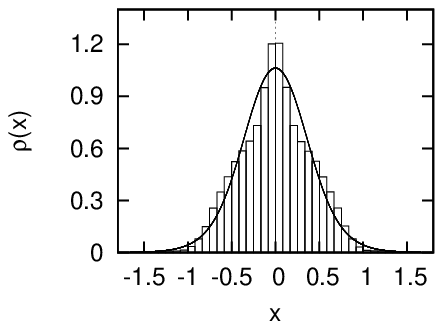}   \hspace{-5mm}
& \includegraphics[width=4.7cm]{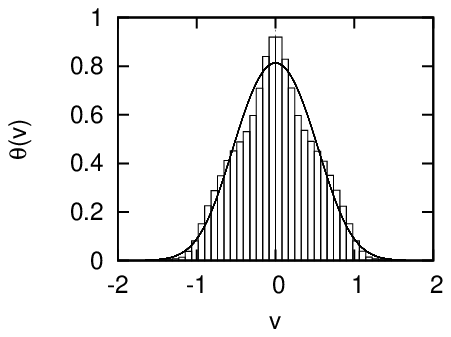}
\end{tabular}
\includegraphics[width=5.5cm]{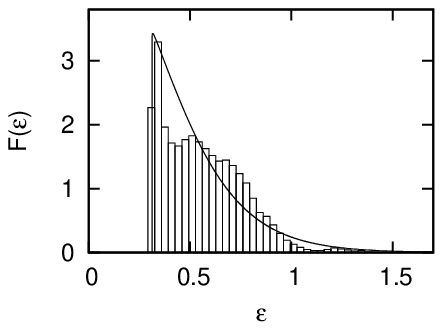}
\caption{The density profile (top left), velocity distribution (top right) and energy distribution (bottom) 
for the QSS obtained starting from the DW1 initial conditions ($\xi_{D}=1.59$). The solid curves
lines are the LB predictions.} 
\label{pdfs_DR1}
\end{center}
\end{figure}

To further explore these findings, and in particular to investigate the relevance of
the initial virial ratio as a parameter, we consider finally a few other ``disjoint"
waterbag  initial conditions as described above. We report results for
the four cases shown in Fig.~\ref{crossing_ic}.  
Each of the initial conditions has been adjusted to have $R_0=1$, and
the values of  the normalized energy are $\xi_D=1.59, 0.58, 0.49$ and 
$0.23$ for DW1 to DW4 respectively. We take in each case a single
realization with $N=10^4$ particles, and calculate a time average
by sampling on $100$ equally spaced time slices in the time
window $[4000,5000]$ (in the time units of our simulation,
which differ in each case from units with $t_c=1$ by a numerical
factor of order unity). Shown in Fig.~\ref{crossing_phi} are the QSS 
obtained as represented in the $(\phi_{11}, \phi_{22})$ plane. In each
case the filled symbol represents the corresponding LB 
predictions. Comparing to the results for SRW and DRW 
initial conditions, the QSS appear in all cases much closer to 
the LB predictions.   This is confirmed by inspection of the
distribution functions, which are shown for DW2 in 
Fig.~\ref{pdfs_DR3}  and for DR1 in Fig.~ \ref{pdfs_DR1}.
For the former case the results are as close to the LB predictions 
as for the SRW and DRW cases which gave best agreement with LB,
with the small deviation being visible again in the energy distribution 
but very difficult to discern in $\rho(x)$ or $\theta(v)$. The results for
the cases DR3 and DR4 are similar. For DR1, on the other hand, the deviation
from LB is much more marked, and we see in the energy space
that this deviation is associated to the formation of a (in this
case very small) core. Very much in line with the results
for SRW and DRW initial conditions,  the agreement with LB thus 
deteriorates as one goes away from the degenerate
limit.

In summary these results confirm the conclusion drawn from the
analysis of the SRW and DRW waterbags: the LB predictions are always
reasonably good --- and excellent for the spatial and velocity distributions --
for (waterbag) initial conditions with low $\xi_D$, but even at higher
values good agreement can be obtained in cases characterized by
an initial virial ratio of order unity. Further deviation from LB is always characterized
by the appearance of a core-halo type structure.


\section{Comparison with theoretical proposals beyond LB:
direct analysis of phase space density} 
\label{comparison}

Let us consider how well two recent proposals in the literature 
can account for the properties of the QSS we observe:

\begin{itemize}

\item  Yamaguchi \cite{yamaguchi_2008}  studies the SGS model for 
SRW initial conditions, and notes (as was remarked also in early studies
\cite{goldstein+cuperman+lecar_1969,   lecar+cohen_1971}) that the breakdown 
of LB theory is  associated with the apperance of a core-halo structure. He proposes 
a phenomenological adaptation of LB theory which he uses to fit the resultant core, 
in which the LB theory is applied only to the mass and energy associated
to the core. In practice this means that one parameter is measured
{\it a posteriori} from the observed QSS.  

\item Levin et al. in a series of works on other models --- plasmas
\cite{levin+pakter+teles_2008},  3D gravity \cite{levin+pakter+rizzato_2008},
2D gravity \cite{teles_etal_2010} and, most recently, the HMF model
\cite{pakter+levin_2010} --- have proposed that, when LB theory
breaks down, QSS correspond to the phase space density: 
\begin{equation}
f(x,v) = f_0 [\Theta (e_F-e)+ \chi\, \Theta (e- e_F) \Theta (e_h-e)] \,. 
\label{levin-ansatz}
\end{equation}
As in the case of  \cite{yamaguchi_2008}, this involves the 
addition of one parameter compared to LB theory. However, 
a physical explanation is proposed for the core-halo form of 
(\ref{levin-ansatz}),  and a prediction for this additional 
parameter is derived from the initial conditions:  An
analysis of particle dynamics in the coherent oscillating
field associated with the relaxation shows that there
are dynamical resonances which allow particles to
gain energy, with $e_h$ corresponding to the maximal 
energy which can be attained in this way. Assuming that
resonance effect is ``shut off" only by the upper bound 
on the phase space density imposed by the collisionless
dynamics, the ansatz (\ref{levin-ansatz}) is the
simplest one possible for the QSS which will result.
\end{itemize}

\begin{figure}[!h]
\begin{center}
\begin{tabular}{cc}
\hspace{-6mm}
\includegraphics[width=4.7cm]{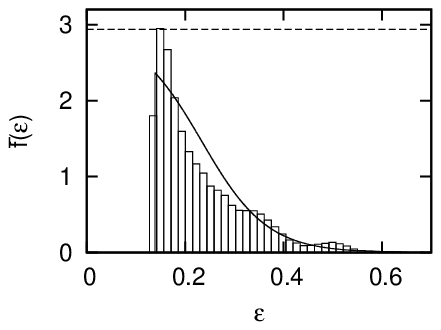} \hspace{-5mm} &
\includegraphics[width=4.7cm]{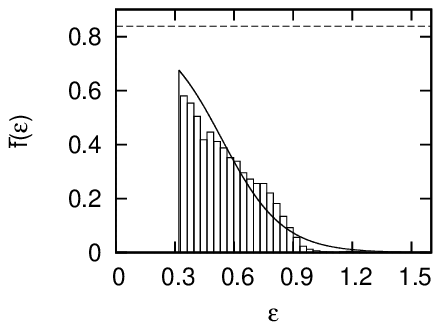} \\   
\hspace{-6mm}
\includegraphics[width=4.7cm]{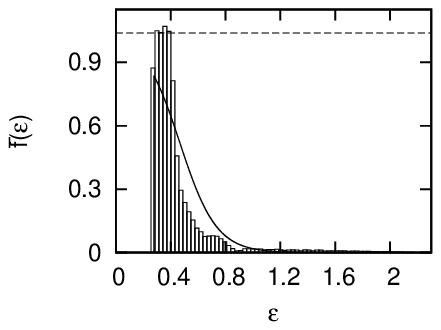} \hspace{-5mm} & 
\includegraphics[width=4.7cm]{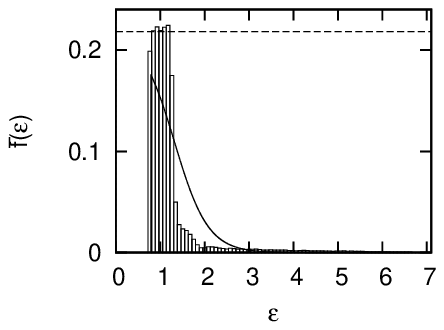} \\
\hspace{-6mm}
\includegraphics[width=4.7cm]{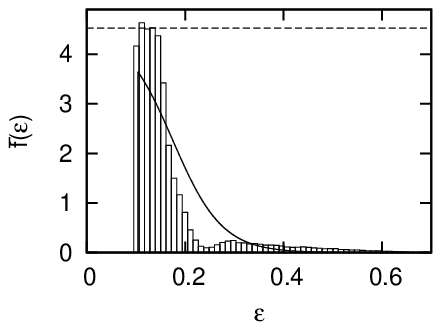} \hspace{-5mm} & 
\includegraphics[width=4.7cm]{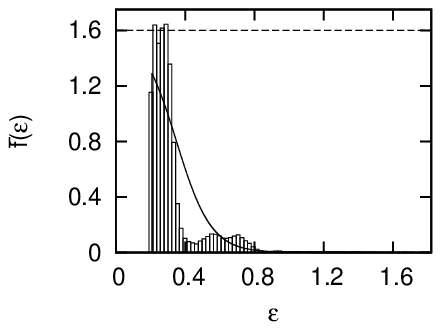} \\
\hspace{-6mm}
\includegraphics[width=4.7cm]{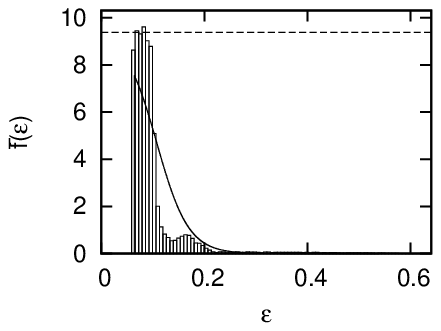} \hspace{-5mm} & 
\includegraphics[width=4.7cm]{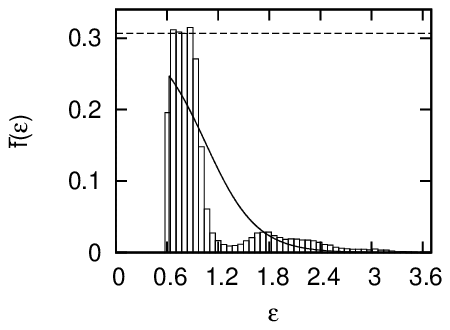} 
\end{tabular}
\caption{Phase-space density as a function of energy $\bar{f}(\epsilon)$ for eight representative
cases of DRW  initial conditions with $\xi_D=0.56$ (corresponding to $R_{0}=0.1$ for SRW). 
The two upper panels correspond to the two cases for which the distribution functions are
shown in Figs.  \ref{noncorehalo1} and \ref{noncorehalo2} where the QSS is close to LB. 
The dashed horizontal line indicates the initial phase space density, $f_{0}$, while
the continuous lines correspond to the LB prediction.}
\label{fexv_lb}
\end{center}
\end{figure}

\begin{figure}[!h]
\begin{center}
\begin{tabular}{cc}
\hspace{-6mm}
\includegraphics[width=4.7cm]{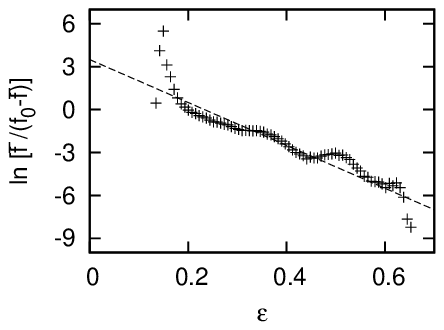} \hspace{-5mm} &
\includegraphics[width=4.7cm]{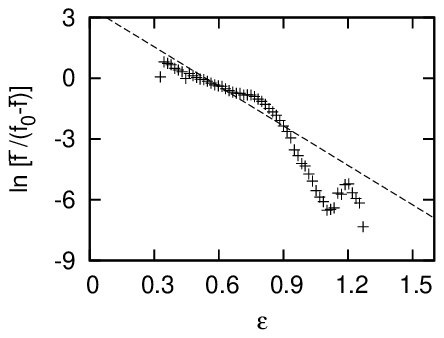} \\   
\hspace{-6mm}
\includegraphics[width=4.7cm]{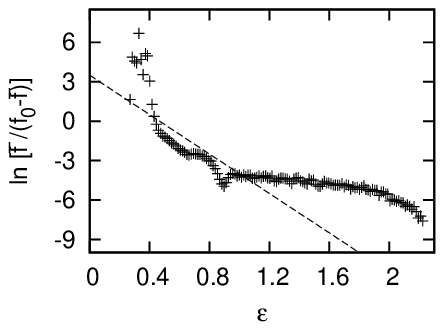} \hspace{-5mm} & 
\includegraphics[width=4.7cm]{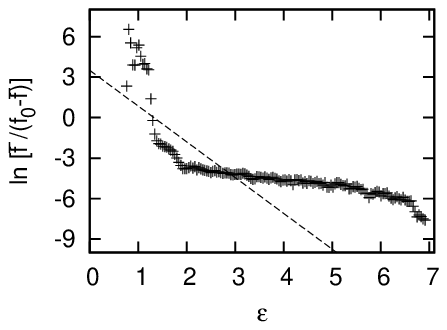} \\
\hspace{-6mm}
\includegraphics[width=4.7cm]{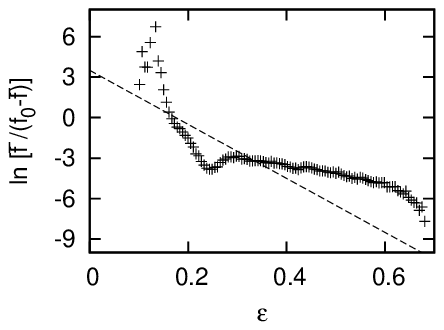} \hspace{-5mm} & 
\includegraphics[width=4.7cm]{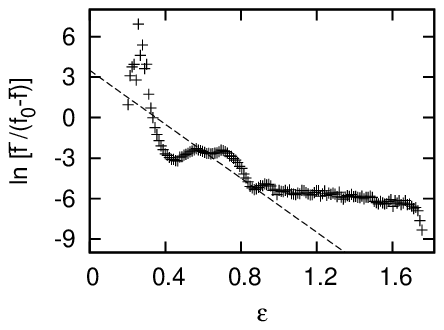} \\
\hspace{-6mm}
\includegraphics[width=4.7cm]{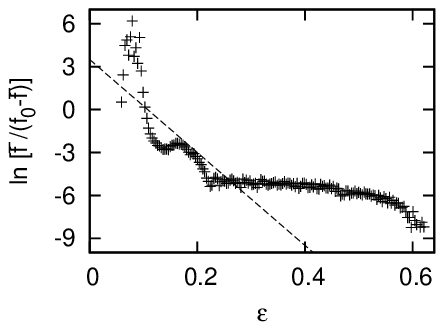} \hspace{-5mm} & 
\includegraphics[width=4.7cm]{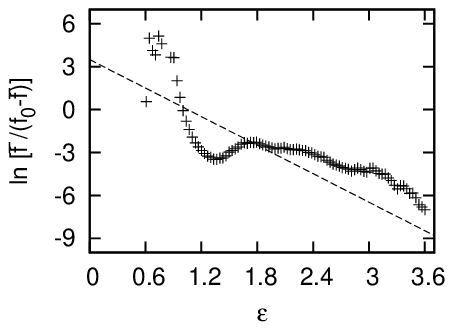} 
\end{tabular}
\caption{Exactly the same data as in the previous figure, but now with the logarithm
of the absolute value of $\bar{f}/(f_0 -\bar{f})$ plotted as a function of particle
energy $\epsilon$. The dashed lines represent the predictions of LB theory (which
become straight lines of slope $\beta$ in this representation). 
}
\label{fexv_lb_log}
\end{center}
\end{figure}

To evaluate the validity of these approaches in this model, we consider directly
the measured phase space density,  $\bar{f}(\epsilon)$, as a function of particle 
energy. To do so we measure the (averaged) values of the potential $\phi(x)$ and $a(\varphi)$ 
in the QSS, and then use (\ref{density-states-shifted})  to calculate the phase space 
density $D(\epsilon)$.  Shown in Fig.~\ref{fexv_lb} are the results for eight chosen cases 
from the DRW initial conditions with $\xi_D=0.56$ considered in Sec.~ \ref{numedouble}
above.  In Fig.~\ref{fexv_lb_log} a plot of exactly the same data is given, but now 
displaying the logarithm of the absolute value of $\bar{f}/(f_0 -\bar{f})$ as a function 
of $\epsilon$ (which in LB theory gives a straight line with slope $\beta$).
Our choice is representative of the whole batch of initial conditions, in
that 1) most QSS have a clear core-halo structure, and 2) those that do not
agree reasonably well with the LB prediction. Indeed the two configurations in 
the uppermost panel of Fig.~\ref{fexv_lb} are the same two cases for which the full 
distribution functions were shown in Figs. ~\ref{noncorehalo1} and \ref{noncorehalo2}. 

In Fig.~\ref{fexv_lb} a vertical line indicates the initial phase phase 
density $f_0$, so that it is clear that whenever a core appears it is
indeed {\it degenerate}. While the measured phase space distributions
are clearly more structured than (\ref{levin-ansatz}), in most cases 
this simple ansatz gives a reasonably good  fit (i.e. about as close to 
the phase space density as the LB profile is to the observed one in the 
cases where it has been considered to work well above).
The slightly greater structuration of the phase space density compared
to the ansatz of (\ref{levin-ansatz}) can also be seen in Fig.~\ref{fexv_lb_log},
which shows in particular that the diffuse halo, when present, although close to flat,  
appears clearly more consistent with a Maxwell-Boltzmann form (i.e. the
non-degenerate limit of LB theory). In this respect we note that
Levin et al. have not tested their ansatz directly against the phase
space density, but have used it to derive predictions for
$\rho(x)$ and $\theta(v)$ which have been compared with those observed.
As we have seen in comparing numerical results with LB predictions
above, these quantities typically wash out structure in energy 
space and make it difficult to see discrepancies which are localized
in this space. We note further that our finding that it is initial conditions 
with $R_0 \approx 1$ which suppress the core-halo formation --- 
and lead to QSS in reasonable agreement with LB --- 
appears completely coherent with the mechanism described
by Levin et al.: when the system starts close to virial equilibrium, the 
relaxation is typically indeed much ``gentler", simply because the 
system does not undergo the large contractions  and expansions which  
result necessarily if there is a large imbalance between the
initial potential and kinetic energy. It is precisely such 
macroscopic oscillations of the system which drive the
resonances analyzed by Levin et al.. 

We consider finally comparison of our results  with an analytical treatment 
of collisionless relaxation developed in  \cite{chavanis_KTql_2008} (see also
further references therein). This work develops, under certain approximations and
hypotheses, a kinetic equation for collisionless relaxation --- similar to the
Lenard-Balescu equation for collisional relaxation --- with a term 
describing relaxation towards the LB equilibrium. One feature of this term is 
that it involves an effective space and time dependent diffusion coefficient, which is proportional
to the product $\bar{f}(f_0 -\bar{f})$. Thus the theory suggests that relaxation 
should be expected to be most inefficient when $\bar{f}$ is close to 
degenerate ($\bar{f}\simeq f_0$) or very
small ($\bar{f} \simeq 0$). In regions of energy where relaxation is more complete, the
distribution is expected to approach the LB form, but with values of the
parameters $\beta$ and $\mu$ different from those in the global LB
equilbrium. Our results in Figs.~\ref{fexv_lb}  and \ref{fexv_lb_log}
do appear to be quite consistent with these qualitative predictions: indeed 
this theory would appear to account for why it is core-halo type states, whose 
dynamical origin is explained  by Levin et al., which do not relax to 
the (global) LB equilibrium. 
In all cases the results in Fig.~\ref{fexv_lb_log}  show a region where the halo
distribution is very consistent with a Maxwell-Boltzmann form with an inverse temperature 
lower than that  of the global LB prediction (dashed line). This can be interpreted,
following  \cite{chavanis_KTql_2008},  as a ``mixing region''  where the 
(in this case, non-degenerate) LB distribution applies locally, while the 
deviation from the (local)  LB form at higher and lower energies is 
considered as due to incompleteness of relaxation in these regions. 
Further the fact that the observed distributions are, compared to the extrapolated 
straight  line (``local" LB) fit in the``mixing region'',  sensibly higher at lower energies
and lower at the highest energies is also in apparent
agreement with the kinetic theory described in  \cite{chavanis_KTql_2008} .

\section{Conclusion and discussion} 
\label{conclusion}

We summarize now our principal conclusions:

\begin{itemize}

\item  {\bf Attainment of QSS in the SGS model:}  
In all cases we have considered QSS do appear to be attained, but 
the time-scales for relaxation to them can vary very considerably. 
In some cases rotating  ``holes" formed in the phase space density during 
the initial phase of violent relaxation ($\sim 10^{2} t_c$)  survive for 
quite a long time, disappearing only on times scales of order
($\sim 10^{3} t_c$). As on these latter time scales the system shows
no apparent tendency to relax towards its thermodynamic equilibrium,
we conclude that this is simply a manifestation of slow collisional
relaxation, and does not imply that QSS are not attained  as argued
by  \cite{mineau+rouet+feix_1990, rouet+feix}. A fuller study of 
the possible $N$ dependence of such relaxation would be useful
to establish this conclusion more firmly (but would be numerically
challenging). 

\item {\bf LB theory in the SGS model:} As was clear already from early studies 
\cite{goldstein+cuperman+lecar_1969,   lecar+cohen_1971}, and confirmed by more recent ones
such as \cite{yamaguchi_2008},  LB theory is not an adequate theory
for understanding fully, or even approximately,  the properties of QSS arising
from violent relaxation in the SGS model {\it for arbitrary initial conditions}. 
However, it is by no means an irrelevant theory to understanding these QSS.
Our study of a quite broad range of initial conditions shows
that  the space of QSS in this model divides quite neatly into two: those for which LB 
works to quite a good approximation, and those for which the phase space
density is characterized by a degenerate core, taking a form generally
quite close to the simple ansatz  (\ref{levin-ansatz}) proposed by Levin et al. 
The initial conditions in the former class are either close to the degenerate limit,
or in other cases characterized by initial virial ratio of order unity.
These conditions are precisely those, in line with what has
been described by Levin et al.,  which suppress resonances
which otherwise act very efficiently to produce the degenerate
core-halo structure.

\item  {\bf Accuracy of predictions of QSS in the SGS model:}  
While, as just described, the QSS which result from violent relaxation
divide into those which are close to the LB theory, on the one hand, or
to the ansatz of Levin et al., on the other,  the accuracy of the 
associated predictions is at best approximate: {\it in no case} do we
see a perfect agreement with either LB theory or the ansatz  (\ref{levin-ansatz}).
We underline that in this respect the spatial distribution of mass $\rho(x)$
and velocity distribution $\theta(v)$ are rather poor tools for diagnosing
the agreement between observations and theory, as they wash
out deviations which are most pronounced in energy space.
We have also noted the apparent coherence of  our results
with the qualitative predictions of the kinetic theory approach 
described in  \cite{chavanis_KTql_2008}.

\end{itemize}

Numerous results in the literature on various other models (see
references in introduction) for specific ranges of initial conditions 
suggest that these latter two conclusions, and probably the first 
also, might apply much more generally to long-range systems.  
Further detailed investigation of such models, and in particular of broader
classes of waterbag initial conditions like those considered here,
or, for example,  ``multi-level" waterbag initial conditions
would be required to establish if this is the case. 

For the SGS it would be interesting to apply the analysis described 
by Levin et al. to determine a prediction of the form (\ref{levin-ansatz}) for 
different initial conditions, and see how well it does in approaching the observed
QSS. In this respect it is interesting perhaps to note that, at given
value of $\xi_D$ this is a one parameter family of solutions, so that
it predicts QSS lying on a curve in the ($\phi_{11}, \phi_{22}$) plane.
In Fig.~\ref{double01}, we see that the QSS obtained from 
the two parameter family of initial conditions at a fixed 
$\xi_D=0.56$ do approximately collapse onto a curve. We would
expect the degree to which the simple ansatz (\ref{levin-ansatz}) can fit
the QSS to be well characterized by determining the prediction 
it gives in this plane.

Of particular interest is of course the original context of 3D self-gravitating 
systems, to which the initial study of  \cite{levin+pakter+rizzato_2008}  for 
SRW suggests these conclusions may indeed apply. As mentioned,
however, the results reported have been based, in this case, on examination
of the density profile $\rho(x)$ alone, while the energy distribution is probably
a finer diagnostic tool as we have seen here.  In forthcoming 
work we will study this case, and discuss the possible relevance
of our findings in the astrophysical context. In this respect we
note one of the reasons why LB theory has not played --- at
least for what concerns it detailed predictions --- a role in
astrophysics is that these predictions depend on unobservable
initial phase space densities. In contrast the prediction of
a degenerate core in many cases would give a simple
link between observations and initial conditions, which
may be of practical relevance notably in constraining
the parameters in theories of structure formation in the
universe. 

The simulations were carried out in large part at the Centre de Calcul
of the Institut de Physique Nucl\'eaire et Physique des Particules.
We are particularly grateful to Laurent Le Guillou for advice and help
on use of these computing resources. We thank B. Marcos for
useful discussions, and P.H. Chavanis for many helpful remarks, and 
in particular for suggesting the plot in Fig.~\ref{fexv_lb_log}. 

\appendix
\section{Determination of $\beta$ and $\mu$} \label{findmubeta}
\begin{figure}[!h]
\begin{center}
\includegraphics[width=8cm]{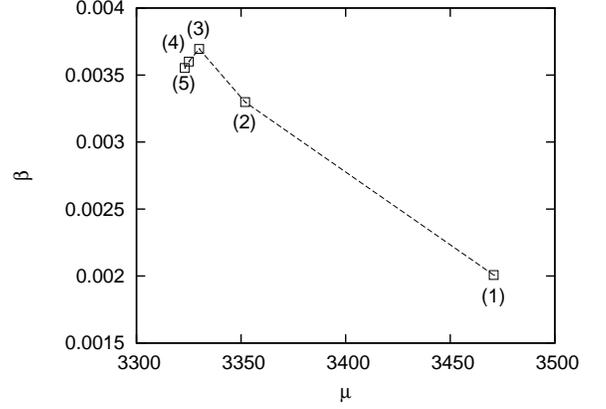} 
\caption{The values of $\beta$ and $\mu$ obtained in the successive steps of our
iterative numerical calculation, for a typical case. The units here are those used in our
numerical calculation ($M=N$, $g=1$ and $L_0=N$), different to those in which
our results in the main text are given.} \label{convergelb}
\end{center}
\end{figure}

In general $\beta$ and $\mu$ cannot be calculated analytically, 
so we solve for them numerically as follows.
The mass normalization (\ref{mconstrain}) condition is 
\begin{equation}
M = \int_{-\infty}^{\infty}\int_{-\infty}^{\infty} \bar{f}(x,v) dxdv. \nonumber
\end{equation}
Integrating over $v$, and changing the coordinate $x$ to $\varphi (x)$
just as in (\ref{eqveloprofile1}), we obtain
\begin{equation}
M = 4\int_{0}^{\infty}\int_{0}^{\infty} \frac{\bar{f} (\varphi,v)}{a(\varphi)} dvd\varphi.  \label{normm}
\end{equation}
The total energy constraint (\ref{econstrain}), i.e.,  
\begin{equation}
E = \int_{-\infty}^{\infty}\int_{-\infty}^{\infty} (\frac{v^{2}}{2}+\frac{\varphi(x)}{2}) \bar{f}(x,v) dxdv,  \nonumber
\end{equation}
can likewise be rewritten as
\begin{eqnarray}
E &=& 2\int_{0}^{\infty}\int_{0}^{\infty}  
\frac{v^{2}\bar{f} (\varphi,v)}{a(\varphi)} dvd\varphi 
\nonumber \\ & & 
+2\int_{0}^{\infty}\int_{0}^{\infty}  
\frac{\varphi \bar{f} (\varphi,v)}{a(\varphi)} dvd\varphi  \label{norme} \\
&=& T+U, \nonumber
\end{eqnarray}
where $T$ is total kinetic energy and $U$ is total potential energy. 
We can then use the virialization condition, $2T = U$, to obtain
\begin{eqnarray}
E = 6\int_{0}^{\infty}\int_{0}^{\infty} 
\frac{v^{2}\bar{f} (\varphi,v)}{a(\varphi)} dvd\varphi .\label{norme_virial}  
\end{eqnarray}
The determination of the parameters $\beta$ and $\mu$ in the LB solution
(\ref{lbequation}) can then be cast as the problem of finding the solutions
of the equations
\begin{eqnarray}
F(\beta,\mu) &=& 0 \nonumber \\
G(\beta,\mu) &=& 0. \nonumber
\end{eqnarray}
where 
\begin{eqnarray}
F(\beta,\mu) &=& M-4\int_{0}^{\infty}\int_{0}^{\infty} \frac{f_{0}}{1+e^{\beta(\frac{v^{2}}{2}+\varphi-\mu)}}
\cdot\frac{1}{a(\varphi)} dvd\varphi \nonumber \\ & & \label{fm} \\
G(\beta,\mu) &=& E-6\int_{0}^{\infty}\int_{0}^{\infty}\frac{v^{2}\cdot f_{0}}{1+e^{\beta(\frac{v^{2}}{2}+\varphi-\mu)}}
\cdot\frac{1}{a(\varphi)} dvd\varphi . \nonumber \\ & & \label{ge}
\end{eqnarray}
Following a standard method we write the matrix equation
\begin{equation}
\left( \begin{array}{c} dF(\beta,\mu) \\ \\ dG(\beta,\mu) \end{array} \right) = \left( \begin{array}{ccc} 
\frac{\partial F}{\partial\beta} & & \frac{\partial F}{\partial\mu} \\ & & \\ \frac{\partial G}{\partial\beta} & & 
\frac{\partial G}{\partial\mu} \end{array} \right) \left( \begin{array}{c} d\beta \\ \\ d\mu \end{array} \right) 
\label{jacobi1}
\end{equation}
where $dF$ and $dG$ denote the infinitesimal changes of $F$ and $G$ when $(\beta,\mu)$ 
change to $(\beta+d\beta,\mu+d\mu)$, we start by guessing a pair of $(\beta,\mu)$ and 
then determining the new $(\beta ',\mu ')=(\beta + \Delta\beta,\mu 
+ \Delta\mu)$ using
\begin{equation}
\left( \begin{array}{c} \Delta\beta \\ \\ \Delta\mu \end{array} \right) = \left( \begin{array}{ccc} 
\frac{\partial F}{\partial\beta} & & \frac{\partial F}{\partial\mu} \\ & & \\ \frac{\partial G}{\partial\beta} & & 
\frac{\partial G}{\partial\mu} \end{array} \right)^{-1} \left( \begin{array}{c} \Delta F(\beta,\mu) \\ \\ 
\Delta G(\beta,\mu) \end{array} \right) \label{jacobi2}
\end{equation}
where 
$(\Delta F,\Delta G)=(-F(\beta,\mu),-G(\beta,\mu))$. 
We then iterate until  $\Delta F$ and $\Delta G$ converge to $0$.
With a reasonable guess for the starting values of $\beta$ and $\mu$,
good convergence is attained within a few iterations, as illustrated
in Fig.~ \ref{convergelb} for a typical case.
\section{Degenerate limit of LB theory} \label{degenerate}
For completeness we reproduce here the analytic results of 
\cite{lecar+cohen_1971} (see also \cite{chavanis_PRE2004})
for the degenerate limit  of the LB distribution function (\ref{lbequation}). 
This corresponds to $\beta\rightarrow\infty$, in which
\begin{equation}
f(x,v) = \left\{ \begin{array}{ll} f_{0}, & \epsilon (x,v) < \mu \\
                                 0, & \textrm{otherwise.}  
               \end{array} \right. \label{degenerate1}
\end{equation}
The density profile is then 
\begin{eqnarray}
\rho (\varphi) &=& 2\int_{0}^{\sqrt{2(\mu-\varphi)}} f_{0} dv \nonumber \\
&=& 2\sqrt{2}f_{0}[\mu-\varphi]^{\frac{1}{2}} \label{degenerate2}
\end{eqnarray}
and therefore, using (\ref{a3}),
\begin{eqnarray}
a(\varphi) 
&=& 4\cdot 2^{\frac{1}{4}}\cdot (\frac{gf_{0}}{3})^{\frac{1}{2}}[\mu^{\frac{3}{2}}-(\mu-\varphi)^{\frac{3}{2}}]^{\frac{1}{2}}.
\label{degenerate3}
\end{eqnarray}
The mass normalization then yields
\begin{eqnarray}
M &=& 2\int_{0}^{\mu}\frac{\rho(\varphi)}{a(\varphi)}d\varphi \nonumber \\
&=& \frac{4\sqrt{2}f_{0}}{\sqrt{\frac{16\sqrt{2}gf_{0}}{3}}}\int_{0}^{\mu}
\frac{(\mu-\varphi)^{\frac{1}{2}}d\varphi}{(\mu^{\frac{3}{2}}-(\mu-\varphi)^{\frac{3}{2}})^{\frac{1}{2}}} \nonumber \\
&=& 2^{\frac{5}{4}}(\frac{f_{0}}{3g})^{\frac{1}{2}}\int_{\varphi=0}^{\varphi=\mu}
\frac{d[\mu^{\frac{3}{2}}-(\mu-\varphi)^{\frac{3}{2}}]}{[\mu^{\frac{3}{2}}-(\mu-\varphi)^{\frac{3}{2}}]^{\frac{1}{2}}}, 
\nonumber
\end{eqnarray}
which can be integrated to give
\begin{equation}
M=2^{\frac{9}{4}}(\frac{f_{0}}{3g})^{\frac{1}{2}}\mu^{\frac{3}{4}}. \label{degenerate5}
\end{equation}
Using the expression (\ref{norme_virial}) for the total energy we have
\begin{equation}
E_D=6f_{0}\int_{0}^{\mu}\int_{0}^{\sqrt{2(\mu-\varphi)}}\frac{v^{2}}{a(\varphi)}dvd\varphi. 
\nonumber
\end{equation}
Integration first over $v$ gives
\begin{eqnarray}
E_D
&=& 2^{\frac{1}{4}}(\frac{3f_{0}}{g})^{\frac{1}{2}}\int_{0}^{\mu}\frac{(\mu-\varphi)^{\frac{3}{2}}}{[\mu^{\frac{3}{2}}-(\mu-\varphi)^{\frac{3}{2}}]^{\frac{1}{2}}}
d\varphi
\nonumber
\end{eqnarray}
and then, on integrating by parts, we obtain
\begin{equation}
E_D=\frac{2^{\frac{13}{4}}}{3}(\frac{f_{0}}{3g})^{\frac{1}{2}}\mu^{\frac{7}{4}}
\int_{0}^{1}(1-\varphi')^{\frac{1}{2}}\varphi'^{-\frac{1}{3}}d\varphi'. \nonumber
\end{equation}
where $\varphi '=(\frac{\mu-\varphi}{\mu})^{\frac{3}{2}}$.
The integral can be expressed as a beta function, and the result
can thus be written
\begin{equation}
E_D=\frac{2^{\frac{13}{4}}}{3}(\frac{f_{0}}{3g})^{\frac{1}{2}}B(\frac{3}{2},\frac{2}{3})\mu^{\frac{7}{4}}. \label{degenerate13}
\end{equation}
It is simple to show from (\ref{norme}) that $\frac{\partial E}{\partial\beta}<0$ in general, tending asymptotically to
$0$ as $\beta \rightarrow \infty$. This is thus indeed the minimal possible energy corresponding to given $M$
and $f_{0}$.

\section{Generation of DRW initial conditions} \label{appen_db}

For the DRW phase space density defined in Sec.~\ref{doublewb}
a direct calculation gives immediately that the initial kinetic energy is
\begin{equation}
T_{0}=f_{0}[\frac{2(x_{2}-x_1)v_{2}^{3}}{3}+\frac{2x_{1}v_{1}^{3}}{3}] \nonumber
\end{equation}
and the initial potential energy 
\begin{eqnarray}
U_{0}&=&4f_{0}^{2}v_{2}^{2}g[\frac{4x_{2}^{3}}{3}+\frac{2x_{1}^{3}}{3}-2x_{2}^{2}x_{1}] \nonumber \\
& &+8f_{0}^{2}v_{1}v_{2}g[x_{1}(x_{2}^{2}-x_{1}^{2})]+\frac{16}{3}f_{0}^{2}v_{1}^{2}gx_{1}^{3}. \nonumber
\end{eqnarray}
These indeed reduce to the corresponding expressions (\ref{t_u}) for the SRW
(when we set  $x_{1}=0$, $x_{1}=x_{2}$ or  $v_{1}=v_{2}$). To generate the specific initial condition 
reported in Sec.~\ref{numedouble}, we do a random
sampling in  in $x_{1}, x_{2}, v_{1}$ and $v_{2}$ at fixed $f_{0}$ , $E$ and $M$ fixed (which implies
that $\xi_D$ is fixed). We then choose configurations with $R_{0}=2T_{0}/U_{0}$ and $\delta$ as various 
as possible. 


\end{document}